\documentclass[RNAAS]{aastex62}

\usepackage{epsfig}
\usepackage{color}
\usepackage{amsmath} 
\usepackage{amssymb} 
\usepackage{enumerate}
\usepackage{float}
\usepackage{dirtytalk}

\shorttitle{Determining the Composition of Jets from Polarization}

\begin{document}

\title{DETERMINING THE COMPOSITION OF RELATIVISTIC JETS FROM POLARIZATION MAPS}

\correspondingauthor{Richard Anantua}
\email{ranantua@cfa.harvard.edu}

\author[0000-0003-3457-7660]{Richard Jude Anantua}
\affiliation{Center for Astrophysics $\vert$ Harvard \& Smithsonian, 60 Garden Street, Cambridge, MA 02138, USA}
\affiliation{Black Hole Initiative at Harvard University, 20 Garden Street, Cambridge, MA 02138, USA}

\author{Razieh Emami}
\affiliation{Center for Astrophysics $\vert$ Harvard \& Smithsonian, 60 Garden Street, Cambridge, MA 02138, USA}

\author{Abraham Loeb}
\affiliation{Center for Astrophysics $\vert$ Harvard \& Smithsonian, 60 Garden Street, Cambridge, MA 02138, USA}
\affiliation{Black Hole Initiative at Harvard University, 20 Garden Street, Cambridge, MA 02138, USA}

\author{
Andrew Chael
}
\affiliation{
Princeton Center for Theoretical Science, Jadwin Hall, Princeton University, Princeton, NJ 08544, USA
}

\begin{abstract}

We present a stationary, axisymmetric, self-similar semi-analytic model of magnetically dominated jet plasma based on force-free regions of a relativistic magnetohydrodynamic simulation. We  use this model to illustrate how the composition of relativistic jet plasma can be determined, with special attention to the example of M87. 
In particular, we compute synthetic Stokes maps 
in $e^-e^+p$ plasmas with various positron-to-proton ratios  
using synchrotron emission models 
scaling the partial pressure of electrons and positrons emitting at the observed frequency to the magnetic pressure, taking into account  Faraday rotation and conversion. 
The lepton-dominated models produce bilaterally asymmetric radio 
intensity profiles with strong linear polarization and Stokes $Q$ and $U$ maps that are bilaterally asymmetric (but strongly 
correlated across the jet axis) and antisymmetric (and sometimes 
anticorrelated), respectively. The hadronic models produce more centrally brightened intensity and polarization maps. Circular polarization provides the cleanest observational tool for distinguishing the plasmas, as it increases outward from the jet core and central axis 
for highly ionic plasma, and is suppressed for  
pair-dominated plasma. We find a measurable degree of circular polarization $V/I$ of $\mathcal{O}(10^{-3})$ for sub-equipartition hadronic jet plasmas. Our stationary model predicts that the intensity-normalized autocorrelation functions of $Q$ 
and $U$ 
increase and decrease with frequency, respectively. On the other hand, the autocorrelation of $V$ 
is less sensitive to the frequency. Multi-band polarimetric observations 
could therefore be used as a novel probe of the composition of jet plasma. 
\end{abstract}  


\section{Introduction}

One of the most fundamental mysteries surrounding relativistic jets in active galactic nuclei (AGN) is the relative proportion of positrons and protons in them \citep{Begelman1984}. Jet regions near the magnetic axis tend to be highly magnetized and are often modeled invoking the Blandford-Znajek (BZ) mechanism \citep{1977MNRAS.179..433B,McKinney2012}, in which BZ jet power scales as the squares of black hole spin and the horizon magnetic flux, and the electric potential drop across field lines is sufficient to produce $e^-e^+$ pairs that emit synchrotron radiation.   
Accretion disks at the base of these jets may be modeled as ion torii providing varying degrees of pressure support for jets based on their thickness and density. An intermediate disk-jet region surrounding the BZ jet may be powered by hydrodynamic winds carrying ions from the disk in the Blandford-Payne (BP) mechanism \citep{1982MNRAS.199..883B}. The ultimate composition of the low density relativistic jet plasma depends sensitively on how efficiently ions are screened by the barrier of the rotating magnetic field lines around it. 

Since over a century ago when Heber Curtis observed a curious ``straight ray" 
in the galaxy M87 (also known now to host the radio source 3C 274)
, 
observations of relativistic jets have increased in spatiotemporal resolution. 
Recent observations of AGN jets, such as the minute-scale variable 3C 279 in $>100$ MeV gamma rays \citep{FermiLAT2016} and the limb brightened M87 jet and counterjet at 43 GHz \citep{Walker2018} observed with the Very Long Baseline Array (VLBA) have challenged our understanding of particle acceleration of these most energetic multiwavelength astrophysical sources \citep{pierre2017}. Circular polarization measurements in 3C 279 attaining values as high as 1$\%$ indicate particle distribution functions so dominated by the low ($\gamma<<100$) end that Faraday rotation would preclude strong linear polarization if the jet were $e^-p$, thus suggesting positron dominance for that source \citep{Wardle1998}. Shocked ionic plasma can reasonably account for M87 boundary layers in cm intensity profiles, but the degree to which Doppler boosting and other jet particles affect the final images are uncertain. At smaller scales towards the jet base, other AGN components such as molecular clouds and the accretion disk may obscure the jet emission. 

The advancement of mm very long baseline interferometry (VLBI) with the intercontinental network of baselines forming the Event Horizon Telescope (EHT) enabled observations of M87 at 230 GHz with resolution down to the black hole horizon \citep{EHT2019}. This allows direct comparison with general relativistic magnetohydrodynamic (GRMHD) simulations, such as High-Accuracy Relativistic Magnetohydrodynamics (HARM) \citep{McKinney2003}, which evolve plasma rest mass density $\rho$, energy density $u$, velocity 4-vector $v^\mu=\gamma(c,\vec{v})$ and magnetic 4-vector\footnote{Fluid frame magnetic 4-vector $b^\mu$ is related to lab frame $
B_\nu=\gamma(\vec{v}\cdot\vec{B},\vec{B}-\vec{v}\times\vec{E})$ via $b^\mu=\frac{h^\mu_\nu B^\nu}{u^t}$, where projection tensor $h_\nu^\mu=(v^\mu v_\nu+\delta^\mu_\nu)$.} $b^\mu$ near black holes. However, there is still substantial uncertainty in the radiation physics that light up these simulations, and how this physics changes at increasing radii from the black hole. To help free ourselves from the constraints of the spatiotemporal dynamical range of the simulations, we step back and use a semi-analytic model inspired by them. We endow this model with self-similarity, which enables us to model processes close to the black hole, such as the conversion of magnetic energy into pairs, as well as processes along the entire jet length, such as the buildup of current into magnetic towers \citep{Fowler2019}.
    
In the following, we compare jet plasmas across a range 
of possible compositions, 
from a pure electron-positron 
to predominately electron-proton 
plasma. The goal is to find decisive observational tests to distinguish these 
cases. To this end, we seek some observables that are sensitive to the charge-to-mass ratio of the particles. In particular, we expect the observed polarization to be dependent on plasma composition due to Faraday conversion and rotation effects \citep{Marrone2007}, 
and thus compute linear and circular polarization of an $e^{-}e^{+}$ plasma and compare autocorrelations of these maps with those of maps with an $e^{-}p$ plasma.
\footnote{Note that the energy of particles contributing to  polarized synchrotron emission is proportional to $m^{-3/2}$ \citep{1968ApL.....2....1R}, where $m$ refers to the radiating particle's mass. Therefore, we would expect that the total emission from the $(e^{-}e^{+})$ plasma to be higher than the emission from the $(e^{-}p)$ as protons are 1,836 times heavier than electrons and do not significantly contribute to the synchrotron emission (unless 
their velocity distribution is greatly skewed towards $c$). 
There is also suppression of ionic power due to the to the $t_ \mathrm{Syn,i}/t_\mathrm{Syn,e} \sim (m_p/m_e)^4$ slower synchrotron cooling time for ions \citep{1968ApL.....2....1R}. Thus, the direct contributions of 
ions
to synchrotron radiation is neglected in our analysis.}


The contribution of the electrons is degenerate with that of the positrons and their respective number densities appear additively in the radiative transfer functions relating to emission and linear polarization. 
The degeneracy is broken by the consideration of circular polarization, which cancels out for a symmetric $e^+e^-$ plasma, but does not for an $e^-p$ plasma. In what follows, we use protons and ions interchangeably, as we do not expect observationally viable signatures to have a metallicity dependence due to the subdominant cosmic abundances of elements heavier than hydrogen. Jet composition has been addressed in previous analytic models including that of \citet{Park2010}, who find that ion number density and magnetic field strength have degenerate effects on the change in electric field phase angle through jet plasma; however, the semi-analytic approach presented here self-consistently models emission as well as propagation. 

The paper is organized as follows. In Sec. \ref{synch}
we present the synchrotron formalism. In Sec. 
\ref{Radiative-Transfer} we present a review of basic transfer equations for the synchrotron emission.
In Sec. \ref{EmissionModeling}, we specify the parametric emission prescription for turning GRMHD variables into observable radiation from a given distribution of electrons, positrons and protons. 
In  Sec. \ref{SemiAnalyticModel} we present our stationary, self-similar, axisymmetric semi-analytic model 
which we analyze through synthetic observations-- Stokes maps, degree of polarization maps and autocorrelations--  in Sec.  \ref{Results}. 
We conclude in Sec. \ref{conc} with a summary of our main results and their implications.

\section{Synchrotron Formalism}
\label{synch}

We start by pedagogically constructing polarized emission $j'_{\nu\Omega}$ and absorption coefficeints $\chi'_{\nu\Omega}$ from the particle distribution function $N'_{e \Omega'}$ per unit solid angle (where primes denote fluid comoving quantities). Polarized synchrotron radiation is characterized by emission and absorption coefficients that depend linearly on the partial pressure 
of electrons and positrons emitting near the observed frequency. 
The radiative transfer functions derived after specifying comoving electron+positron number density per unit solid angle \citep{Anantua2016}, $N'_{e\Omega'}$, will enable us to implement emission models simply by specifying phenomenologically motivated partial pressures. Faraday conversion of circular to linear polarization also has a similar 
dependence, though  Faraday rotation of linearly polarized light does not.

We start with coefficients of the radiative transfer equation in the emitting particle frame, then convert them into observer frame quantities by making use of the Lorentz invariance of $\nu\chi_\nu$ and $j_\nu/\nu^2$ and noting the Doppler factor:
\begin{equation} 
\mathcal{D}=\frac{1}{\gamma \left(1-\frac{\vec{v}}{c}\cdot \hat{n}\right)}=\frac{\nu}{\nu'},
\end{equation}
where $\nu$ is the observed frequency,  $\hat{n}$ is the direction along the line of sight (l.o.s.) (cf. Fig. \ref{ObserverPlane}), and primed quantities denote the particle frame. Then we solve the radiative transfer equation by l.o.s. integration of observer frame quantities. 
In the next subsection, we decompose emission and absorption onto the distant observer ($XY$-plane) using the basis set by the two independent polarization directions of an incoming photon. 
In what follows, we assume emitting particles are isotropic in the emitting plasma rest frame.

\subsection{Power Law Electrons and Positrons}

First consider a power law distribution of electron and positron energies:
\begin{equation}
    N'_{e\Omega'}(\gamma')=K_{e\Omega'}'\gamma'^{-p},\ \ \  K'_{e\Omega' }=\frac{1}{4\pi}(n'_{e^-}+n'_{e^+})(p-1). 
\end{equation}
where $\gamma’$ is the Lorentz factor of the emitting particles and $n'_e$ is their combined number density. The partial pressure contribution from electrons and positions emitting at the observed frequency can be written as:
\begin{equation}
    \tilde{P}_e=\frac{4\pi}{3}\gamma'^2N'_{e\Omega'}(\gamma')m_ec^2=\frac{4\pi}{3}K_{e\Omega'}'m_ec^2\gamma'^{2-p}.
\end{equation}   
We assume that the particles follow the fluid with $v_{e-}^\mu\approx v_{e+}^\mu\approx v^\mu$ 
\citep{Ressler2015,Ressler2017}, 
although different particle species may have slight relative speeds 
\citep{Bourouaine2013}.
We distinguish the thermal Lorentz factor, $\gamma$, from the bulk Lorentz factor, $\Gamma$.





\subsection{Observing a Relativistic Emitter 
} 


Adding an observer to 
the formalism, it is natural to express the synchrotron radiation theory formulae in terms of quantities parallel to the line of sight and quantities along the polarization plane. In Figure \ref{PolarizationAxes}, the polarization plane is on the sky and $\zeta$ is a coordinate along the line of sight in the direction $\hat{n}$. This introduces a viewing angle dependence through:
\begin{figure}
\begin{center} 
\includegraphics[height=275pt,width=350pt]{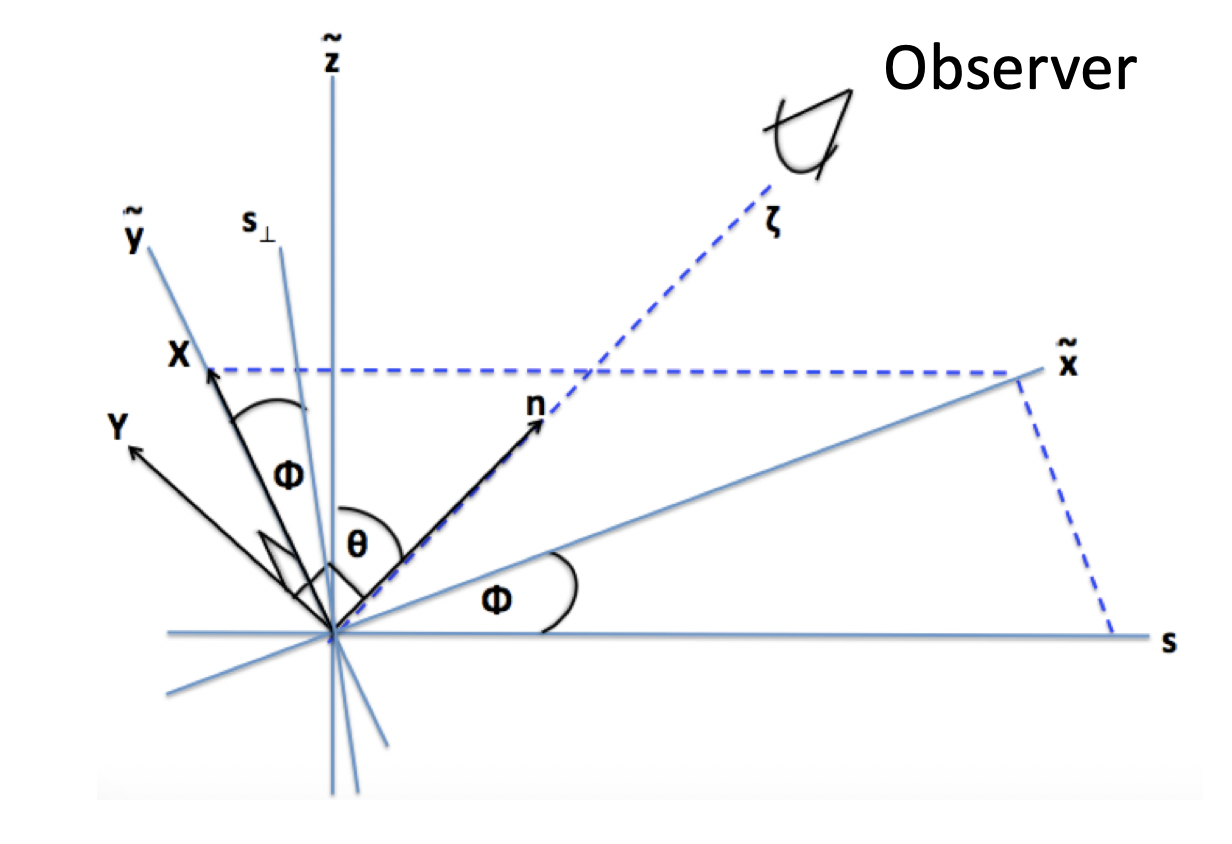}
\caption[Polarization Axes on the Sky]{Polarization axes $X$ and $Y$ on the sky superposed onto galaxy Cartesian coordinates $\{\tilde{x},\tilde{y},\tilde{z}\}$ and galaxy cylindrical polar coordinates $\{s,\phi,\tilde{z}\}$. The displacement from the observer plane is parameterized by $\zeta$ and the inclination angle is $\theta$. A jet propagating to length $L$ along $\tilde{z}$ in the galaxy has projected length $L\cos\theta$ along the observer $Y$ axis.}\label{ObserverPlane}
\end{center}
\label{PolarizationAxes}
\end{figure}

\begin{equation}
B_e\equiv\mathcal{D}^{-1}|\vec{B}_\perp'|=\mathcal{D}^{-1}|\vec{B}'\times\hat{n}|,   
\end{equation}
Particle energy is linked to emitted photon frequency by:
\begin{equation}
\nu_c=\frac{3eB_\perp'}{4\pi m_e c}\gamma'^2
\end{equation}
where $\gamma'=\gamma$ and $\mathcal{D}=\mathcal{D}^{'-1}$ due to their respective $v$ dependences. 
Though emission observed at any frequency comes from a range of energies, it is convenient to replace the observing frequency with twice the critical frequency in the analytical calculations below.

\subsection{Emission and Absorption}

The comoving frame emissivity  and absorption coefficients can be derived from the power per unit frequency radiated by an electron gyrating around a magnetic field \citep{Westfold1959}:
\begin{equation}
p'_{\nu'1,2}(x)=\frac{\sqrt{3}\mu_0e^3cB'_\perp}{4\pi m_e}\left[\frac{x}{2}\left(\pm K_{2/3}(x)+\int_x^\infty K_{5/3}(x')dx'\right)\right],\ \ \mathrm{\ }x\equiv\frac{\nu'}{\nu_c'},
\label{PolarizedSynchrotronP}
\end{equation}
where $K_q$ is a modified Bessel function of imaginary argument and 1 and 2 refer to two independent polarization directions and $\nu_c=\frac{3}{4\pi}\frac{eB_\perp}{m_ec}\gamma^2$. Then the comoving emissivity components for a power law electron
distribution are:
\begin{equation}
\begin{tabular}{cr}
$j'_{\nu'\Omega'1,2}=K'_{\Omega'}\left(\frac{\nu_c'}{\gamma'^2}\right)^\frac{p-1}{2}\int_0^\infty dx 2x^\frac{p-3}{2}p'_{\nu'1,2}(x)$\\ $=\frac{\mu_0e^3cB'_\perp K'_\Omega}{8\sqrt{3}\pi m_e}\left(\frac{3eB_\perp'}{2\pi\nu'm_e}\right)^\frac{p-1}{2}\frac{\Gamma\left(\frac{3p-1}{12}\right)\Gamma\left(\frac{3p+7}{12}\right)}{p+1}\{\frac{3p+5}{2},1\}$\\
$=\frac{1}{4\sqrt{3}}\tilde{P}_e\left(\frac{r_e\nu'}{c}\right)\left(\frac{3eB'_\perp}{2\pi \nu'm_e}\right)^\frac{3}{2}\frac{\Gamma\left(\frac{3p-1}{12}\right)\Gamma\left(\frac{3p+7}{12}\right)}{p+1}\{\frac{3p+5}{2},1\},$\\
\end{tabular}\label{SynchEmisOpticThickPol}
\end{equation}
and the comoving absorption is:
\begin{equation}
\begin{tabular}{cr}
$\chi'_{\nu 1,2}=-\frac{1}{m_e\nu'^2}\int d\gamma'\gamma'^2\frac{d}{d\gamma'}\left(\frac{N'_{\gamma',\Omega'}}{\gamma'^2}\right)p'_{\nu'1,2}(\gamma)
$\\$=\frac{\sqrt{3}(p+2)\mu_0e^3cB'_\perp K'_\Omega}{16\pi m_e^2\nu'^2}\left(\frac{3eB_\perp'}{4\pi\nu'm_e}\right)^\frac{p}{2}\left(\pm K_{2/3}(x)+\int_x^\infty dx'K_{5/3}(x')\right)$\\ 
$=\frac{\mu_0 e^3cB'_\perp K'_{\Omega'}}{8\pi\sqrt{3}m_e^2\nu'^2}\left(\frac{3eB'_\perp}{2\pi\nu'm_e}\right)^{p/2}\Gamma\left(\frac{3p+2}{12}\right)\Gamma\left(\frac{3p+10}{12}\right)\{\frac{3p+8}{2},1\}$\\
$=\frac{1}{4\sqrt{3}}\left(\frac{\tilde{P}_e}{m_3\nu'^2}\right)\left(\frac{r_e\nu'}{c}\right)\left(\frac{3eB'_\perp}{2\pi\nu' m_e}\right)^2\Gamma\left(\frac{3p+2}{12}\right)\left(\frac{3p+10}{12}\right)\{\frac{3p+8}{2},1\}$,\\
\end{tabular}\label{SynchAbsorbOpticThickPol}
\end{equation}
where $\Gamma(\cdot)$ is the gamma function  and $r_e=\frac{e^2}{m_ec^2}$ is the classical electron radius.

The observer frame components of the emissivity and absorption coefficient associated with the independent polarization directions are found using the Lorentz transformation properties of emission, absorption, magnetic field and frequency \citep{Anantua2016}:
\begin{equation}
\begin{tabular}{cr}
$j_{\nu\Omega 1,2}(\nu,\hat{n},\vec{B})=\mathcal{D}^4P_S\frac{r_e\nu}{4\sqrt{3}c}\left(\frac{3eB_e}{2\pi \nu m_e}\right)^\frac{3}{2}\frac{\Gamma\left(\frac{3p-1}{12}\right)\Gamma\left(\frac{3p+7}{12}\right)}{p+1}\{\frac{3p+5}{2},1\}$,\\
$\chi_{\nu 1,2}(\nu,\hat{n},\vec{B})=\mathcal{D}^4P_S\frac{r_e}{4\nu\sqrt{3}c}\left(\frac{3eB_e}{2\pi \nu m_e}\right)^2\Gamma\left(\frac{3p+2}{12}\right)\Gamma\left(\frac{3p+10}{12}\right)\{\frac{3p+8}{2},1\}$,
\end{tabular}\label{LabEmisAndAbsorp}
\end{equation}
where 
there is a strong viewing angle dependence through $\mathcal{D}=(1-\frac{|\vec{v}|}{c}\cos\theta)^{-1}$. The total emission and absorption coefficients per solid angle are then:
\begin{equation} 
\begin{tabular}{cr} 
$j_{\nu\Omega}\equiv j_{\nu\Omega 1} + j_{\nu\Omega 2} = \frac{3(p+1)}{3p+7} (j_{\nu\Omega 1} - j_{\nu\Omega 2})$,\\
$\chi_{\nu\Omega}\equiv \frac{1}{2}(\chi_{\nu\Omega 1} + \chi_{\nu\Omega 2})=\frac{3(p+2)}{3p+10}(\chi_{\nu\Omega 1} - \chi_{\nu\Omega 2})$.
\end{tabular}\label{GalaxyEmisAndAbsorp}
\end{equation}
We see the emission is completely specified by $P_S$. Models based on this formalism \citep{Anantua2016,
2017JPhCS.840a2023B,
Anantua2018} have used  $P_S=\tilde{P}_e$; however to compare plasmas with varying $n_{e^+}/n_p$ we may generalize to $P_S=f(\tilde{P}_e,\tilde{P}_i)$.

This methodology 
casts synchrotron 
emission and absorption functions in terms of the pressure $\tilde{P}_e$ due to electrons or positrons emitting at the observed frequency and the effective magnetic field $B_e=\mathcal{D}^{-1}|\vec{B}\times\hat{n}|$ for observer direction $\hat{n}$ (note we perform slow light radiative transfer that does not approximate the speed of light as infinite). This formslism serves as the basis for 1- or 2-parameter models that show promising agreement with observations of AGN, 
e.g., M87 \citep{Anantua2018
}.

\section{Full Polarized Radiative Transfer}
\label{Radiative-Transfer}

The 
full polarised radiative transfer equation in the fluid frame 
reads \citep{2016MNRAS.462..115D}:
\begin{align}\label{PolarizedRad}
\frac{d}{d s} \begin{pmatrix} I'  \\ Q'  \\ U' \\ V'
  \end{pmatrix}  = \begin{pmatrix} j'_I  \\ j'_Q  \\ j'_U \\ j'_V\end{pmatrix}  - \begin{pmatrix} \chi'_I & \chi'_Q  & \chi'_U &  \chi'_V \\
 \chi'_Q &  \chi'_I  &  \rho'_V  & \rho'_U \\
   \chi'_U  & -\rho'_V &  \chi'_I  & \rho'_Q   \\
     \chi'_V  & -\rho'_U & -\rho'_Q  & \chi'_I 
\end{pmatrix} \begin{pmatrix} I'  \\ Q'  \\ U' \\ V'
  \end{pmatrix},
\end{align}
where $(I', Q', U', V')$ refer to the Stokes parameters; $j'_I,j'_Q,j'_U,j'_V$ represent the polarized emissivities; $\chi'_I,\chi'_Q,\chi'_U,\chi'_V$ are the absorption coefficients; and $\rho'_V$ and  $\rho'_Q, \rho'_U$  refer to one Faraday rotation and two Faraday conversion coefficients, respectively. 

We wish to solve Eq. (\ref{PolarizedRad}) for different plasma compositions. We may simplify the analysis by aligning the magnetic field with the Stokes $U'$ parameter to get $j'_U = X'_U = \rho'_U = 0 $ \citep{2016MNRAS.462..115D}. The most generic solution can then be achieved by performing a transformation which changes the polarization basis. 
In the following, we first compute the emissivity, absorption, Faraday rotation and conversion functions in the $U'=0$ basis. We then change the polarization basis and present the form of the transformation matrix to compute the remaining transfer parameters, i.e. $j'_U, \chi'_U$ and $\rho'_U$. 

\subsection{Polarized emission, absorption and Faraday coefficients in the $U' = 0$ basis}\label{ComovingRTFunctions}

Here we compute emissivity, absorption and Faraday coefficients in $U' = 0$ basis. We consider a plasma made of electrons, protons and positrons. Due to the mass hierarchy, we ignore the direct contribution of protons to the radiative transfer equations and just keep them to guarantee the overall charge neutrality of the plasma. 

We also generalize our power law distribution function to electrons and positrons whose energies are truncated at some low and high Lorentz factors:

\begin{equation}
\label{distributionF}
N'_{e\Omega' \gamma} = \frac{1}{4 \pi} N'_{e\gamma} =  
\begin{cases}
     \left(\frac{p-1}{4 \pi} \right) \left(n'_{e^{-}} + n'_{e^{+}} \right) \left( \gamma'^{1-p}_\mathrm{min} - \gamma'^{1-p}_\mathrm{max}  \right)^{-1} \gamma'^{-p} \equiv K'_{e \Omega \gamma} \gamma'^{-p}
	&, ~~~ \gamma'_\mathrm{min} \leq \gamma' \leq \gamma'_\mathrm{max} \\
	~~~~~0 &, ~~~ \mathrm{otherwise}. 
\end{cases}
\end{equation}
where 
$\gamma_\mathrm{min}$ and $\gamma_\mathrm{max}$ refer to the minimum and maximum particle Lorentz factors, respectively. 

Upon computing emissivity, absorption and Faraday coefficients for an $e^{-}e^{+}$ plasma, 
we compare the results at each step with similar expressions in the literature.  
Since locally, the direction of motion of  electrons and positrons around the magnetic field are in opposite senses, 
we adopt overall signs for their coefficients that are sensitive to the projection of the magnetic field vector along the wavevector. As shown below, this affects $j'_V$, $\alpha'_V$ and $\rho'_V$. 

We label quantities in the $U' =0$ basis 
with a sub-index $0$.

\subsubsection{Polarized Emissivity}

In describing the polarized emissivity for an $e^{-}e^{+}$ plasma, we generalize the formalism described by  \citet{2016MNRAS.462..115D} to include positrons. 
The emissivity function for a power law distribution of particles is given by \citep{2016MNRAS.462..115D}: 
\begin{equation}
\label{emissivity}
j'^{
IJ
} = \int_0^{\infty} d \gamma' N'_{\gamma} \eta^{
IJ
},  
\end{equation}
where $\eta'^{
IJ
}$ refers to the 
4x4 
vacuum emissivity 
matrix 
given by \citet{Melrose1971} and is defined in Eq. (A1) of \citet{2016MNRAS.462..115D}.

We also directly compare our final results with \citet{2019JCAP...01..035E}:
\begin{equation}
j' = \begin{pmatrix} j'_I  \\ j'_Q  \\ j'_U \\ j'_V
  \end{pmatrix}
= j'_0 \left(n'_{e^-} + n'_{e^+} \right)
B_{\perp}^{'(p +1)/2} \nu^{' \frac{1-p}{2}}
\begin{pmatrix} 1  \\ q'  \\ u' \\ v'
\end{pmatrix},
\end{equation} 
where we have:
\begin{align}
 j'_{0I} =& \frac{e^2}{c} \left( \frac{e}{2 \pi m_e c} \right)^{\frac{p+1}{2}}  \left(n'_{e^{-}} + n'_{e^{+}} \right) B_{\perp}^{'(p +1)/2} \nu^{'\frac{1-p}{2}}
\left( \frac{3^{\frac{p}{2}} (p-1)} 
{2(p+1)} \right) 
  \nonumber\\
 &~~~~~~~~~ \times
 \frac{\Gamma{(\frac{3p-1}{12})} \Gamma{(\frac{3p+19}{12})} 
}{(\gamma_\mathrm{min}^{'1-p} - \gamma_\mathrm{max}^{'1-p})}, \\
j'_{0Q}  =& \frac{p+1}{p+7/3} j'_{0I} ,  \\
 j'_{0V} =& \frac{e^2}{c} \left(\frac{e}{2 \pi m_e c} \right)^{\frac{p+2}{2}}  \left(n'_{e^{-}} - n'_{e^{+}} \right) B_{\perp}^{'(p +1)/2} \nu^{'\frac{1-p}{2}}
\left( \frac{B'_{\parallel
}}{\sqrt{\nu' B'_{\perp}}} \right)
  \nonumber\\
 &~~~~~~~~~ \times 
 \left( \frac{3^{\frac{p-1}{2}} (p-1)(p+2)} 
{2p} \right) 
 \frac{\Gamma{(\frac{3p+4}{12})} \Gamma{(\frac{3p+8}{12})} 
}{(\gamma_\mathrm{min}^{'1-p} - \gamma_\mathrm{max}^{'1-p})}, 
\label{j0IQV}
\end{align}
where $\nu'$ refers to the comoving frequency. Our generalized formalism in the presence of positrons is in agreement with \citet{2019JCAP...01..035E}.
 
\subsubsection{Faraday Rotation Coefficients}
Next, we compute the Faraday rotation coefficients for  an $e^{-}e^{+}$ plasma: 
\begin{align}
\label{Farady-Rotation}
\rho'_{0V} =& \left(\frac{e^3}{\pi m^2_e c^2}\right) \left(n'_{e^{-}} - n'_{e^{+}} \right)  \frac{(p-1)(p+2)}{(p+1)}
\frac{\gamma^{'-(p+1)}_\mathrm{min} \ln{\left(\gamma'_\mathrm{min}\right)} }{(\gamma_\mathrm{min}^{'1-p} - \gamma_\mathrm{max}^{'1-p})} B'_{\parallel} \nu^{'-2},
\end{align}
Our final expression is consistent with Eq. (A9) of \citet{2002ApJ...573..485R} which adopts a negative sign for the contribution of the positrons to $\rho'_{0V}$. 

\subsubsection{Faraday Conversion Coefficients}
Next, we compute the Faraday conversion coefficients for  an $e^{-}e^{+}$ plasma: 
\begin{align}
\label{Faraday-Conversion}
 \rho'_{0Q} =& -\frac{e^4}{(2\pi)^2 (m_{e} c)^3} \left(n'_{e^{-}} + n'_{e^{+}} \right)   B_{\perp}^{'2} \nu^{'-3}  \frac{(p-1)\gamma^{'2-p}_\mathrm{min}}{(\gamma_\mathrm{min}^{'1-p} - \gamma_\mathrm{max}^{'1-p})} 
  \nonumber\\
 &~~~~~~~~~ \times
 \bigg{[} \left(1 -\left(\frac{e B'_{\perp}}{2\pi m_e c \nu'} \gamma^{'2}_\mathrm{min} \right)^{p/2-1}\right)\left(\frac{p}{2} - 1\right)^{-1} \bigg{]},
\end{align}
Our final expression for the Faraday conversion is consistent with Eq. (A8) of 
\citet{2002ApJ...573..485R}, which has a positive sign for the contribution of positron in $\rho'_{0Q}$. 

\subsubsection{Absorption Coefficients}

Finally, we compute the absorption coefficients for an $e^{-}e^{+}$ plasma:
\begin{align}
 \chi'_{0I} =& \frac{e^2}{16 m_{e} c} \left( \frac{e}{2 \pi m_e c} \right)^{\frac{p+2}{2}} 3^{\frac{p-1}{2}}  \left(n'_{e^{-}} + n'_{e^{+}} \right)   B_{\perp}^{'(p +2)/2} \nu^{'-\frac{p}{2}-2}
  \nonumber\\
 &~~~~~~~~~ \times
(p-1)(3p+10)
 \frac{\Gamma{(\frac{3p+10}{12})} \Gamma{(\frac{3p+2}{12})} 
}{(\gamma_\mathrm{min}^{'1-p} - \gamma_\mathrm{max}^{'1-p})}, \\
\chi'_{0Q}  =& \frac{p+2}{p+10/3} \chi'_{0I} ,  \\
 \chi'_{0V} =& \frac{e^2}{4 m_e c} \left(\frac{e}{2 \pi m_e c} \right)^{\frac{p+3}{2}}  \left(n_{e^{-}} - n_{e^{+}} \right) B_{\perp}^{'(p +1)/2} \nu^{'-\frac{p+5}{2}}
B'_{\parallel}
  \nonumber\\
 &~~~~~~~~~ \times 
 \left( \frac{3^{\frac{p}{2}} (p-1)(p+2)(p+3)} 
{(p+1)} \right) 
 \frac{\Gamma{(\frac{3p+7}{12})} \Gamma{(\frac{3p+11}{12})} 
}{(\gamma_\mathrm{min}^{'1-p} - \gamma_\mathrm{max}^{'1-p})}. \\
\label{alpha0IQV}
\end{align}
Again, our final results are based on the natural generalization of the approaches presented in \citet{2016MNRAS.462..115D} and  \citet{1977ApJ...214..522J}.

\subsection{Transforming to Observer $I, Q, U,$ and $V$}
Given the comoving radiative transfer functions presented above, we now turn to the problem of expressing them in a convenient form for distant observers viewing a relativistic jet at frequency $\nu$. 
\subsubsection{Transforming from Comoving to Observer Frame}
 
The comoving quantities from Section \ref{ComovingRTFunctions} can be transformed into observer frame quantities by using the Lorentz invariance of $\nu'^{-2}j'_\nu$, $\nu'\chi'_\nu$ and $\nu'\rho'_\nu$ and the following dependencies on $\mathcal{D}$:
\begin{align}
\label{DopplerDependencies}\nonumber
\frac{\nu}{\nu'}&=\mathcal{D}, \\
\frac{B_\perp}{B'_\perp}&=\mathcal{D}^{-1}, \\ \nonumber
B_{||}&= \Gamma \mathcal{D}^{-1}\left(B'_{||}+\frac{\vec{B}\cdot\vec{v}}{\frac{1}{\Gamma}+\frac{1}{\Gamma+1}}\right). 
\end{align}

\subsubsection{The Orientation $\psi$ of the Linear Polarization Ellipse}

We start with the case of no circular polarization (Stokes $V=0$) to focus on the relation between $Q$ and $U$. The polarized radiative transfer equation for Stokes $I,Q$ and $U$ reads:
\begin{align}\label{PolarizedRadiativeTransferEqsWOpacity}
\frac{d}{ds}\begin{pmatrix} I  \\ Q  \\ U  
  \end{pmatrix} =  - \begin{pmatrix} \chi_I & \chi_Q & \chi_U  \\
  \chi_Q & \chi_I & 0  \\
   \chi_U  & 0 & \chi_I  
  \end{pmatrix}\begin{pmatrix} I  \\ Q  \\ U  
  \end{pmatrix}+\begin{pmatrix} j_1+j_2 \\ (j_1-j_2)\cos 2\psi \\ (j_1-j_2)\sin 2\psi  
  \end{pmatrix},
\end{align}
where $j_1$ and $j_2$ are polarized emissivities associated with the two independent photon polarization directions; the absorption coefficients are:
\begin{equation}
\begin{tabular}{cr}
  $\chi_I=\frac{\chi_1+\chi_2}{2}$,\\
 $\chi_Q=\frac{(\chi_1+\chi_2)\cos 2\psi}{2}$,\\
 $\chi_U=\frac{(\chi_1+\chi_2)\sin 2\psi}{2}$,\\
\end{tabular}\label{eq:Stokes}
\end{equation}
and the orientation angle in the observer $(XY)$ plane is:
\begin{equation}\psi=\tan^{-1}\left(B_X^2+B_Y^2+E_X^2+E_Y^2-2(E_XB_Y-E_Y B_X)\right).
\end{equation}
The polarization plane is spanned by $\{\hat{\epsilon}_1,\hat{\epsilon}_2\}$. Note $\psi$ is also the angle between the $X$-axis and the projection of the magnetic field onto the observer plane.

\subsubsection{Changing Polarization Basis from $U=0$}
 
We now recover the general radiative transfer case in which all terms may be nonzero. Choosing a polarization basis so that $j_U=0=\chi_0$ amounted to rotating the $U$ subspace by $2\psi$. We let:

\begin{equation}
\vec{j_0} = \begin{pmatrix} j_{0I}  \\ j_{0Q}  \\ 0 \\ j_{0V}
  \end{pmatrix},\ \ \  \vec{s_0} = \begin{pmatrix} I_0  \\ Q_0  \\ U_0 \\ V_0
  \end{pmatrix} = \begin{pmatrix} I_0  \\ Q_0  \\ 0 \\ V_0
  \end{pmatrix},
  \end{equation}
  
  \begin{align}\label{PolarizedRadiativeTransferEqsWOpacity}
A_0 =   \begin{pmatrix} \chi_I & \chi_Q & 0 &   \chi_V\\
  \chi_{0Q} & \chi_{0I} & \rho_{0V}  & 0  \\
   0  & -\rho_{0V} & \chi_{0I} & \rho_{0Q}  \\
     \chi_{0V}  & 0 & -\rho_{0Q} & \chi_{0I}
  \end{pmatrix}.
\end{align}
Then,
\begin{align}\label{PolarizedRadiativeTransferEqsWOpacity}
\frac{d \vec{s_0}}{d\zeta} =  - A_0\vec{s_0}+\vec{j_0}.
\end{align}
is the radiative transfer equation in the $U=0$ basis. 

Now, to transform this basis, let: 
   \begin{align}\label{PolarizedRadiativeTransferEqsWOpacity}
P =   \begin{pmatrix} 1 & 0 & 0 &  0 \\
  0 & \cos(2\psi) & \sin(2\psi)   & 0  \\
   0  & -\sin(2\psi) & \cos(2\psi) & 0  \\
     0  & 0 & 0& 1
  \end{pmatrix}.
\end{align}
We identify:
\begin{align}\label{PolarizedRadiativeTransferEqsWOpacity}
\frac{d (P^{-1}\vec{s_0})}{d\zeta} = P^{-1}\vec{j_0}- PA_0(P^{-1}\vec{s_0})
\end{align}
with transformed radiative transfer equation:
\begin{align}\label{PolarizedRadiativeTransferEqsWOpacity}
\frac{d \vec{s}}{d\zeta} = \vec{j}- A\vec{s},
\end{align}
so we have:
\begin{equation}
\vec{j} = \begin{pmatrix} j_I  \\ j_Q  \\ j_U \\ j_V\end{pmatrix}  = P^{-1}\vec{j_0} = \begin{pmatrix} j_{0I}  \\ j_{0Q}\cos(2\psi)   \\ j_{0Q}\sin(2\psi) \\ j_{0V}
 \end{pmatrix},
 \end{equation}

\begin{align}
A= \begin{pmatrix} \chi_I & \chi_Q  & \chi_U &  \chi_V \\
 \chi_Q &  \chi_I  &  \rho_V  & \rho_U \\
   \chi_U  & -\rho_V &  \chi_I  & \rho_Q   \\
     \chi_V  & -\rho_U & -\rho_Q  & \chi_I 
  \end{pmatrix} = PA_0P^{-1} \\
=   \begin{pmatrix} \chi_{0I} & \chi_{0Q} \cos(2\psi) & -\chi_{0Q}\sin(2\psi) &  \chi_{0V} \\
  \chi_{0Q}\cos(2\psi) &  \chi_{0I} &  \rho_{0V}  & \rho_{0Q}\sin(2\psi)  \\
   -\chi_{0Q}\sin(2\psi)  & -\rho_{0V} &  \chi_{0I} & \rho_{0Q} \cos(2\psi)   \\
     \chi_{0V}  & -\rho_{0Q} \sin(2\psi) & -\rho_{0Q} \cos(2\psi) & \chi_{0I}
 \end{pmatrix}.
\end{align}
Thus, the matrix $P$ manifestly rotates the $QU$ subspace by $2\psi$ and serves to change the basis of the $U=0$ absorption matrix to a generically rotated basis.
 


 
 

\section{Emission Modeling}
\label{EmissionModeling}

There is a plethora of plausible models representing jet emission phenomenology. Assuming jet emission is sourced by  $e^-e^+$ plasma, \citet{Anantua2018} have employed a current density model with intensity dominated by a central jet ``spine" \citep{Hawley2006,Moscibrodzka2016} surrounded by return currents, and a shear model accentuating boundary layers. Also, a simple emission model we refer to as the constant electron-positron beta model (cf. the beta model of \citep{Anantua2018}), where:
\begin{equation}
    \beta=\frac{P_g}{P_B},
\end{equation}
converts a fraction of the  electromagnetic energy density to  gas (particle) energy density. We find this model to be naturally generalizeable to mixed leptonic/hadronic plasmas, as an increase in ion number density is readily represented by a proportional reduction in the emitting electron partial pressure. We base our investigation of the effects of plasma composition on jet Stokes maps on this class of models.

\subsection{Constant $\beta_e$ Model}

We start with the constant electron-positron beta ($\beta_e$) model, where we equate the partial pressure $\tilde{P}_e$ due to a relativistic electrons-positrons gas to a constant fraction of the local magnetic pressure 
$P_B=\frac{B^2}{2\mu_0}$
: 
\begin{equation}
\tilde{P}_e=\beta_{e0}P_B.
\end{equation}
This model is a linear scaling of $e^-e^+$ gas pressure to local magnetic pressure with proportionality constant $\beta_{e0}$, though other scalings such as power laws have been devised, e.g., the magnetic bias model \citep{Anantua2018}.

Equipartition of particle and magnetic energy densities, which may be brought about by magnetic reconnection, holds for $\beta_{e0}\sim 1$. We expect the highly magnetized inner jet regions that we focus upon will be sub-equipartition. As $b$ drops with distance and the jet entrains particles (at a few 100$M$ in our simulation, where $M\equiv GM_\mathrm{BH}/c^2$), the jet becomes mass loaded and particle dominated. To accommodate all of this physics, we construct an emission model for electrons, positrons and protons to use in our semi-analytic model.

\subsection{Ionic Distribution Function} 
 

We now formally incorporate emission from a population of ions that  have a power law distribution into our polarized radiative transfer framework, though we expect their contribution to emission to be small. For simplicity, we think of them as protons:
\begin{equation}
N_{i\gamma'}'=K_i'\gamma'^{-p_i}, ~~~\gamma'_\mathrm{min} \leq \gamma' \leq \gamma'_\mathrm{max}  
\end{equation}
This could however be generalized to other species as long as we respect the overall charge neutrality of the plasma.

\subsubsection{Ion Reduced Constant $\beta_e$ Beta Model}

Generalizing the $e^-e^+$ plasma Constant Electron Beta Model to plasmas for which $n=n_{e^-}+n_{e^+}+n_i=2n_{e^-}$, we adopt:
\begin{align}
 \tilde{P}_e&= \frac{n-n_i}{n}\beta_{e0}P_B,\\
 P_i& = \frac{n_i}{n}P_g=\frac{n_i}{n}(\Gamma_\mathrm{E.o.S.}-1)u_g,\\ 
 \tilde{P}_i& = \left(\frac{m_e}{m_p}\right)^{4+3(p-1)/2}P_i\approx 0, \\ \tilde{P}_S&=\tilde{P}_e+\tilde{P}_i \approx \tilde{P}_e.
\end{align}
where equation of state parameter $\Gamma_\mathrm{E.o.S.}$ ranges from 4/3 for relativistic particles to 5/3 for non-relativistic particles. The motivation for suppressing the contribution from ion pressure is, again, that protons emitting at the observed frequency are $(m_p/m_e)^{3/2}$ times as energetic as the corresponding electrons and that they cool $t_\mathrm{Syn,i}/t_\mathrm{Syn,e} \sim (m_p/m_e)^4$ slower. 
Thus we take the contribution of $\tilde{P}_i$ to our synthetic observations as 0. Also in our modeling, we take $\gamma_\mathrm{min}=10$ to be fiducial, as $\gamma=1$ gives cyclotron radiation and the synchrotron formalism above requires relativistic particles \citep{Rybicki2004}.


\section{Self-Similar Stationary Semi-Analytic Model}
\label{SemiAnalyticModel}

In order to implement our formalism to yield observables such as Stokes maps and electric vectors on the polarization plane, we employ a semi-analytic model abstracted from the low $\beta$, high $\sigma\ (=\mathrm{magnetic\ flux\ density}/\mathrm{particle\ flux\ density})$ jet flow only $\lesssim 100M$ from the black hole in a GRMHD jet simulation with dipole magnetic field and thick, magnetically arrested disk initial conditions described in \citet{McKinney2009,McKinney2012}. Our model is stationary ($\frac{\partial}{\partial t}=0$) and axisymmetric ($\frac{\partial}{\partial \phi}=0$). This well approximates the azimuthally averaged simulation MHD variables in the jet region. 
For the parabolic jet in this model, the self-similarity variable is $\xi=\frac{s^2}{z}$
, where $s$ is cylindrical radius and $z$ altitude in our cylindrical coordinate system (cf. Fig. \ref{PolarizationAxes}).

We express--
in cylindrical coordinates-- the magnetic flux $\Phi=\Phi(\xi(s,z))$, current $I=I(\xi(s,z))$ and fieldline angular velocity $\vec{\Omega}_B(\xi(s,z))$, and relate jet variables to these:
\begin{equation}
\begin{pmatrix} B_s  \\ B_\phi  \\ B_z 
  \end{pmatrix}
=\begin{pmatrix} -\frac{1}{2\pi s}\frac{\partial\Phi}{\partial z}  \\ \frac{I}{2\pi s}  \\ \frac{1}{2\pi s}\frac{\partial\Phi}{\partial s} 
  \end{pmatrix}
  =\begin{pmatrix} \frac{s\Phi'}{2\pi z^2}  \\ \frac{I}{2\pi s}  \\ \frac{\Phi'}{2\pi z}
  \end{pmatrix},
\end{equation}

\begin{equation}
    \vec{E}=\vec{B}\times(\hat{\Omega}\times \vec{r}),
\end{equation}

\begin{equation}
\begin{pmatrix} j_s  \\ j_\phi  \\ j_z 
  \end{pmatrix}
  =\vec{\triangledown}\times\vec{B}
=\begin{pmatrix} -\frac{sI'}{2\pi z^2}  \\ -\frac{s(2z\Phi'+(s^2+4z^2)\Phi'')}{2\pi z^4}  \\ \frac{I'}{\pi z}
  \end{pmatrix},
\end{equation}

\begin{equation}
\rho=\vec{\triangledown}\cdot\vec{E}=-\frac{1}{2\pi z^2}\left(\xi(4z+\xi)\Phi'\Omega_B'+\Omega_B(2(2z+\xi)\Phi'+\xi(4z+\xi)\Phi'')\right),
\end{equation}
using Amp\'ere's law, Ohm's law for ideal MHD, Faraday's law and Gauss' law, respectively. Assuming our jet plasma is force free:
\begin{equation}
\rho\vec{E}+\vec{j}\times\vec{B}=\vec{0},
\end{equation}
the current obeys:
\begin{equation}
4\xi^2\omega_B^2\Phi'\Phi''+4\xi(\Omega_B^2+\xi\Omega_B\Omega_B')\Phi'^2=I'I,
\end{equation}
with the solution:  
\begin{equation}
I=-2\Omega_B\xi\Phi'.
\end{equation}
To get numerical values in this semi-analytic model, we fit  $\Phi(\xi)$ and $\Omega_B(\xi)$ to azimuthally averaged $\vec{B}$, $\vec{v}$, $\rho$ and $u_g$ from the GRMHD simulation. 
We use the following fitting formulas:  
\begin{align}
 \Phi(\xi)& = \tanh(0.3\xi),\\
 \Omega_B(\xi)&= 0.15\exp(-0.3\xi^2)
 .
\end{align}
The corresponding magnetic and velocity fields are: 
\begin{equation}
  \begin{pmatrix} B_s  \\ B_\phi  \\ B_z 
  \end{pmatrix}
=\begin{pmatrix} \frac{\xi}{2\pi s z}\frac{\partial\Phi}{\partial \xi}  \\ \frac{I}{2\pi s}  \\ \frac{\xi}{\pi s^2}\frac{\partial\Phi}{\partial \xi}
  \end{pmatrix},\ \ 
\begin{pmatrix} v_s  \\ v_\phi  \\ v_z 
  \end{pmatrix}
=\begin{pmatrix} \frac{s}{2z}v_z  \\ s\Omega_B(1-v_z)  \\ v_{z0}e^{-0.001s^8/z^4}
  \end{pmatrix},
\end{equation}
where $v_{z0}(z)$ is interpolated from the simulation using $v_{z0}(10)=0.3c,\ v_{z0}(10^{1.7})=0.75c,\ v_{z0}(10^2)=0.95c,\ v_{z0}(10^{3.5})=0.97c$ and $v_{z0}(10^4)=0.99c$. For completeness, we also introduce the total gas pressure fitting formula:
\begin{equation}
    P_g(s)=3.5\times 10^{-7}(s/35)^3\exp{\left(-(s/35)^4\right)},
\end{equation}
though this pressure contribution is neglected in the radiative transfer. The gas pressure rises from 0 at the jet axis to a maximum near the inflow/outflow boundary.


We are now in a position to construct numerical emission, absorption, and Faraday conversion and rotation functions from self-similar variables in this semi-analytic model. Using the above fitting formulae, we can model jet physics on any scale in which the jet plasma is reasonably well represented by force-free GRMHD-- even beyond the $\lesssim 10^5M$ limit of modern jet simulations. Special relativistic effects are determined by the semi-analytic velocity field, the observational frequency and the viewing angle to the jet axis. 
We will adopt a $20^\circ$ viewing angle to emulate M87. Moreover, to convert our code-derived quantities to physical units for M87, we use black hole mass $M_\mathrm{BH,\mathrm{ M87}} = 6.6\times 10^9 M_\odot$ \citep{gebhardt2011} to get length/time conversion $M\equiv r_{g,\mathrm{M87}}=\frac{GM_\mathrm{BH,\mathrm{ M87}} }{c^2}\approx 10^{13}\  \mathrm{m}$
corresponding to light crossing time $\approx 9\mathrm{\ hr}$,
and magnetic flux/field conversion  $\Phi_{H,\mathrm{M87}}=\int_0^{s_\mathrm{max}(z)}2\pi s B_z\approx 10^{26}\  \mathrm{Wb}$ for the jet segments close to the black hole that we consider in the following.

\section{Implementation
}
\label{Results}

Having presented the generic formalism on how to compute the Stokes parameters in a mixed plasma, we now study different aspects of the Stokes maps for different observed frequencies as well as plasma parameters. Since the observations we focus on have been taken at two frequencies 43 \rm{GHz} and  230 \rm{GHz}, we show the behavior of the modeled system at these frequencies. 

\subsection{Intensity Maps
} 

We compare the 
case of $\beta_{e0}=10^{-4}$ and  $n_{e^-}=n/2=n_{e^+}$ to the  
case with $n_{e-}=n/2=n_i$, where $n$ is the total particle number density. Figsures \ref{IntensityMapsLeptonicVsHadronic43} and \ref{IntensityMapsLeptonicVsHadronic230} present the intensity map to power 1/3 at 43 \rm{GHz} and 230 \rm{GHz}, respectively. The plots are transformed by a one-third power $(\cdot)^{1/3}$ to increase contrast for the displayed quantities. The plots show 
a bilaterally asymmetric intensity map in the $e^-e^+$ plasma (left) and 
a lower amplitude, inward core shifted intensity map  core with a limb brightened polarization signature in the $e^-p$ case (right). Note for our synthetic observation, the horizontal axis on which the jet length is projected corresponds to the $Y$ axis of Fig. \ref{ObserverPlane}, and the jet's transverse direction is along $X$.  

We have generated this and subsequent images in this work using the following \software{Mathematica (version 11.3.0, Wolfram et al. 2018)
}
\subsubsection{
Effects of Plasma Composition at $\nu=43$ GHz}
\begin{figure}\nonumber
\begin{align} 
\includegraphics[height=240pt,width=260pt,trim = 6mm 1mm 0mm 1mm]{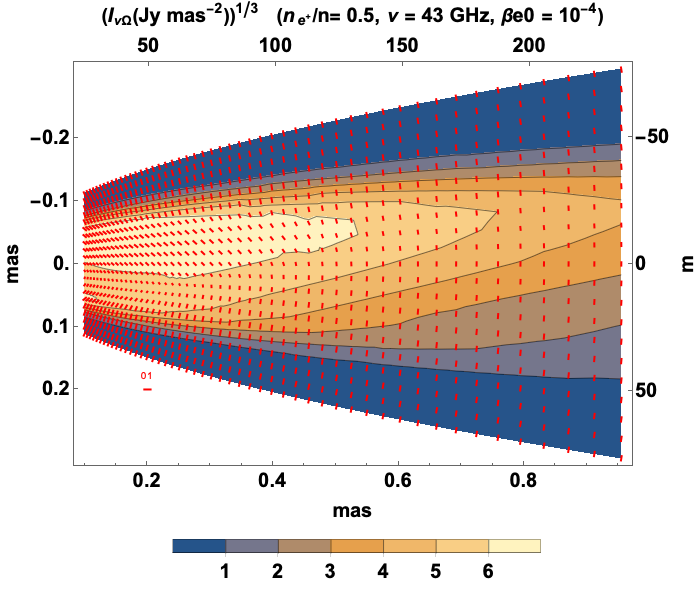}\   &\hspace{1cm}\includegraphics[height=240pt,width=260pt,trim = 6mm 1mm 0mm 1mm]{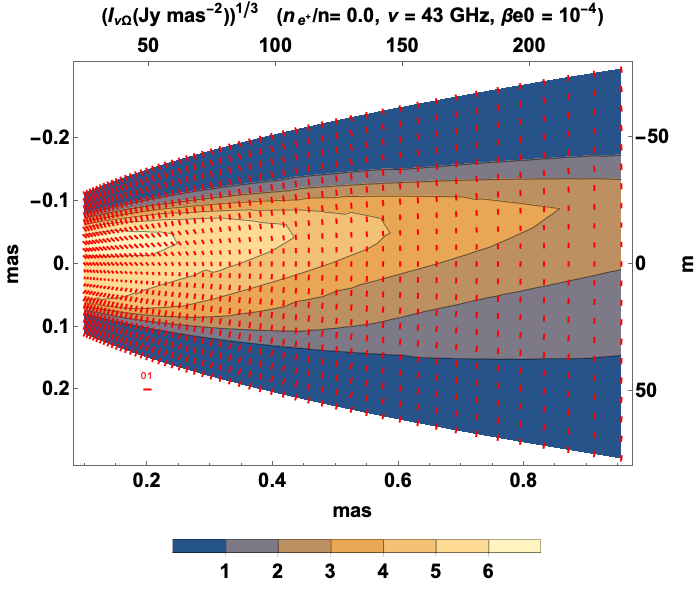}
\end{align} 
\caption{
Images of intensity to the power of 1/3 at 43 \rm{GHz} and $\beta_{e0} = 10^{-4}$ from ray tracing the self-similar stationary semi-analytic model for purely leptonic plasma $n_{e-}=n_{e+}=n/2$ (left) and a maximally hadronic plasma $n_{e-}=n_i=n/2$ (right). Polarization vectors $(P_X,P_Y)$ are also shown (red), oriented at the EVPA.
}\label{IntensityMapsLeptonicVsHadronic43}
\end{figure}

Figure \ref{IntensityMapsLeptonicVsHadronic43} shows that increasing the proton content (decreasing the positron content) diminishes the overall intensity at 43 GHz and shifts the core inward.  
The overall emission from the leptonic plasma is greater, as there are more  emitting particles than when half of the leptons are replaced by protons. The local electric field determining the polarization direction can be written as:
\begin{equation}
(P_X,P_Y)
=\left( \sqrt{\frac{Q^2+U^2+Q\sqrt{Q^2+U^2}}{2I^2}} , \sqrt{\frac{Q^2+U^2-Q\sqrt{Q^2+U^2}}{2I^2}}\mathrm{sgn}(U)\right)
.
\end{equation}

\subsubsection{
Effects of Plasma Composition at $\nu=230$ GHz}
\begin{figure}\nonumber
\begin{align} 
\includegraphics[height=240pt,width=260pt,trim = 6mm 1mm 0mm 1mm]{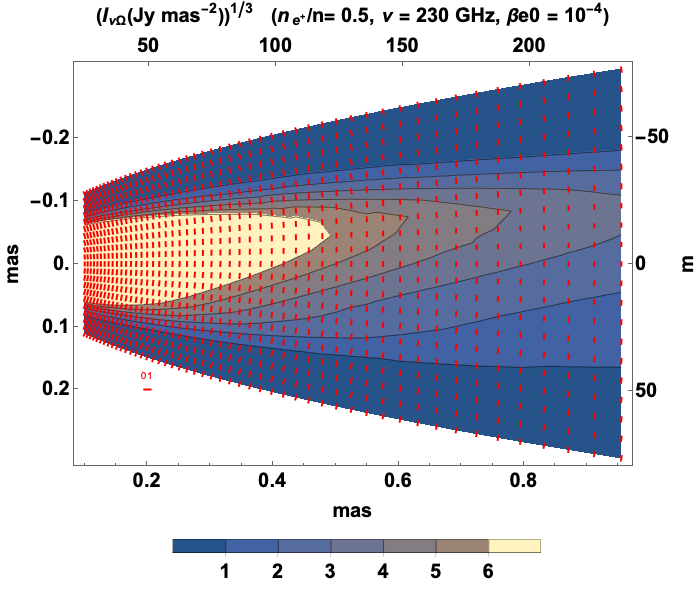}\   &\hspace{1cm}\includegraphics[height=240pt,width=260pt,trim = 6mm 1mm 0mm 1mm]{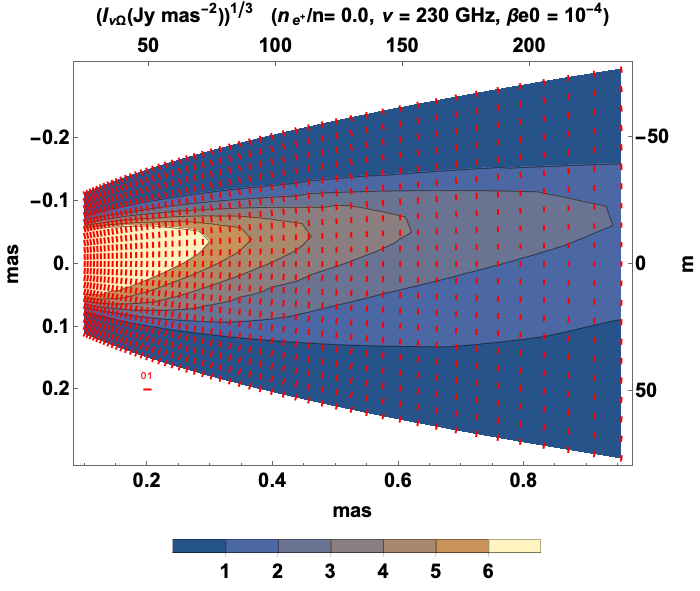}
\end{align} 
\caption{
Images of intensity to the power of 1/3 at 230 \rm{GHz} and $\beta_{e0} = 10^{-4}$ from ray tracing the self-similar stationary semi-analytic model for purely leptonic plasma $n_{e-}=n_{e+}=n/2$ (left) and a maximally hadronic plasma $n_{e-}=n_i=n/2$ (right). Polarization vectors $(P_X,P_Y)$ are also shown (red), oriented at the EVPA.
}\label{IntensityMapsLeptonicVsHadronic230}
\end{figure}

Figure \ref{IntensityMapsLeptonicVsHadronic230} shows the same pattern of intensity decline with increasing ion content, 
now 
shifted further inwards for both compositions due to increased optical depth towards the core
at 230 GHz. The core shift is accompanied by a slight upward shift of the contours within the finite observer plane segments of the maps, slightly enhancing apparent bilateral asymmetry.

\subsection{Stokes Map Symmetry Properties} 
\label{SymProperties}

\begin{figure}[H]\nonumber
\begin{align}
\includegraphics[height=270pt, width=320pt, trim = 6mm 1mm 0mm 1mm]{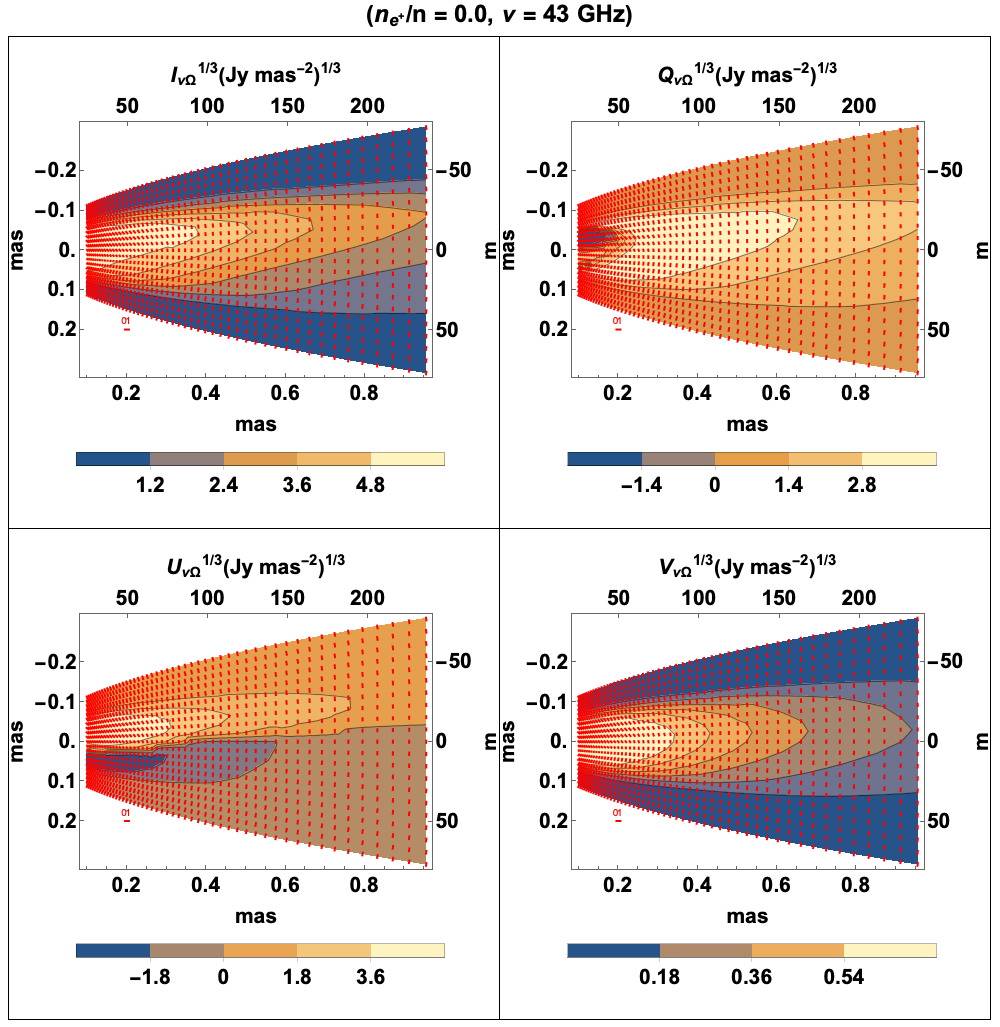} \end{align}  
\begin{align}\nonumber
\includegraphics[height=270pt, width=320pt, trim = 6mm 1mm 0mm 1mm]{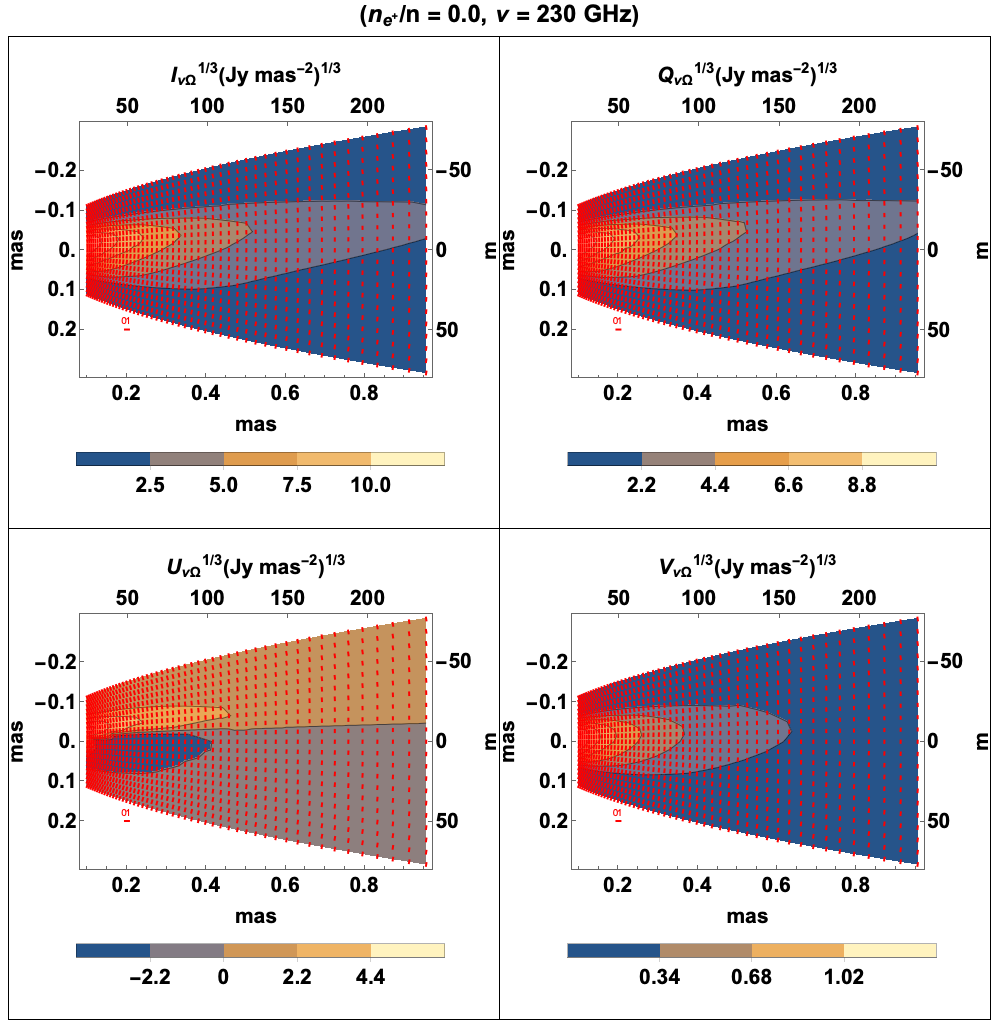}\end{align}
\caption{Stokes map (colors) overlaid with $E$-field vector orientations (red) with $\beta_{e0}=10^{-4}$ at 43 GHz (upper) panel and at 230 GHz (lower) panel. We consider a maximally hadronic plasma with $n_{e+}/n = 0$ and for 
$\gamma_\mathrm{min}=10$, $\gamma_\mathrm{max}=\infty$.
}\label{StokesMapsGamMin10GamMaxInfNu43GHzBetaE010m4neOnnPt0}
\end{figure}

Figures. \ref{StokesMapsGamMin10GamMaxInfNu43GHzBetaE010m4neOnnPt0}, \ref{StokesMapsGamMin10GamMaxInfNu43GHzBetaE010m4neOnnPt25}, \ref{StokesMapsGamMin10GamMaxInfNu43GHzBetaE010m4neOnnPt5}, \ref{StokesMapsGamMin10GamMaxInfNu43GHzBetaE010m2neOnnPt0},
\ref{StokesMapsGamMin10GamMaxInfNu43GHzBetaE010m2neOnnPt25},  and \ref{StokesMapsGamMin10GamMaxInfNu43GHzBetaE010m2neOnnPt5} compare Stokes maps across electron-positron beta  constants $(\beta_{e0}=10^{-4}\ \mathrm{and}\ 10^{-2})$, plasma compositions $(n_i/n=0.0,\ 0.25$ and 0.5) and  frequencies (43 GHz and 230 GHz). We have shown all 4 Stokes parameters (overlaid by the same pattern of electric vectors) for each model parameter choice. The
choice of representing polarization data using electric vector polarization angles (EVPA) overlaid on intensity maps versus displaying all Stokes maps is more than a cosmetic one-- symmetry properties seen in the latter are often concealed in the former.  

We begin our comparison with the lowest $\beta_{e0}$ 
model with lowest ionic content presented in Fig. \ref{StokesMapsGamMin10GamMaxInfNu43GHzBetaE010m4neOnnPt0}, which has the most simply ordered contours: elliptical to parabolic $I$, $Q$ and $V$ and dipolar $Q$. As the composition becomes increasingly leptonic from Fig. \ref{StokesMapsGamMin10GamMaxInfNu43GHzBetaE010m4neOnnPt0} to Fig.  \ref{StokesMapsGamMin10GamMaxInfNu43GHzBetaE010m4neOnnPt25} to Fig.  \ref{StokesMapsGamMin10GamMaxInfNu43GHzBetaE010m4neOnnPt5}, the dipolar pattern of $U$ is mixed into $Q$ and $V$. Increasing the leptonic contribution increases the Stokes $I$ parameter while suppressing the Stokes $V$ parameter. 
As frequency increases from 43 GHz to 230 GHz, the maps tend to have higher overall amplitude emission and $V$ suffers less severe mixing.

Increasing the value of $\beta_{e0}$ from $10^{-4}$ to $10^{-2}$, we see that the higher $\beta_{e0}$ maps in Figs. \ref{StokesMapsGamMin10GamMaxInfNu43GHzBetaE010m2neOnnPt0},
\ref{StokesMapsGamMin10GamMaxInfNu43GHzBetaE010m2neOnnPt25}, and \ref{StokesMapsGamMin10GamMaxInfNu43GHzBetaE010m2neOnnPt5} have higher overall amplitude emission and spatial distribution patterns degenerate with the highly ionic plasmas. We cleave some of this degeneracy by invoking autocorrelations in Section \ref{AutocorrelationAnalysis}.

The most important trend, which can be seen in Figures \ref{StokesMapsGamMin10GamMaxInfNu43GHzBetaE010m4neOnnPt5} and \ref{StokesMapsGamMin10GamMaxInfNu43GHzBetaE010m2neOnnPt5}, is the disappearance of circular polarization, as quantified by the Stokes $V$ parameter for purely symmetric $e^{+}e^{-}$ plasmas ($n_{e^+}/n$ = 0.5). Although, due to the Faraday conversion, the Stokes $V$ parameter does not completely vanish even for the purely symmetric $e^{+}e^{-}$ plasma, its value is far too small to be observable in this case. 
This distinct feature can therefore be used as a fingerprint of proton-deficient jets.
 
 In summary, we see the following main patterns in the Stokes parameters for our model:
 \begin{itemize}
 \item{Bilateral asymmetry in intensity $I$, with maximum brightness towards the core}
 \item{Bilateral asymmetry in one linear polarization, $Q$, which polarization is maximized towards the core}
 \item{Bilateral anti-symmetry in the other linear polarization $U$ (apparently rotated $\pi/4$ from $Q$)}
 \item{Lower amplitude Stokes $V$, which vanishes for vanishing $\beta_{e0}$ or $n_i/n$.}
 \end{itemize}  
 
Another trend is that the symmetry of $Q$ and antisymmetry of $U$ is most apparent for the high frequency, low $\beta_{e0}$, low $n_{e^+}/n$ maps, in which the $Q$ contours are ordered--  from inward to outward-- as 
slightly upward shifted half-ellipses to parallelograms, and the $U$ contours exhibit a simple dipolar pattern symmetric about $Y=0$. At lower frequencies, $Q$ starts exhibiting a dipolar pattern near the core with the negative part on the same side (here upward) as the highest positive intensity regions far away from the hole. At higher $\beta_{e0}$ or $n_{e^+}/n$, $Q$ becomes centrally brightened. As we lower frequency, increase $\beta_{e0}$ or $n_{e^+}/n$ one polarity begins to dominate $U$ on the brightest side (here, upward), distorting the antisymmetry.
 
Observationally, we surmise that $Q$ and $U$ maps in the high frequency, low $\beta_{e0}$ limits in which their contours are most ordered appear 
as $45\%$ phase shifted (cf. \citet{Anantua2018}) versions of each other with opposite symmetric vs. antisymmetric properties. When the order is perturbed by optical depth effects and stronger Doppler beamed emission, the signature is less apparent.
 \newpage

\subsubsection{
Stokes Map Frequency Comparison for $n_{e^+}/n=0.0$,  $\beta_{e0}=10^{-4}$}

\subsubsection{
Stokes Map Frequency Comparison for $n_{e^+}/n=0.25$,  $\beta_{e0}=10^{-4}$ }
\begin{figure}[H]\nonumber
\begin{align}
\includegraphics[height=270pt,width=320pt,trim = 6mm 1mm 0mm 1mm]{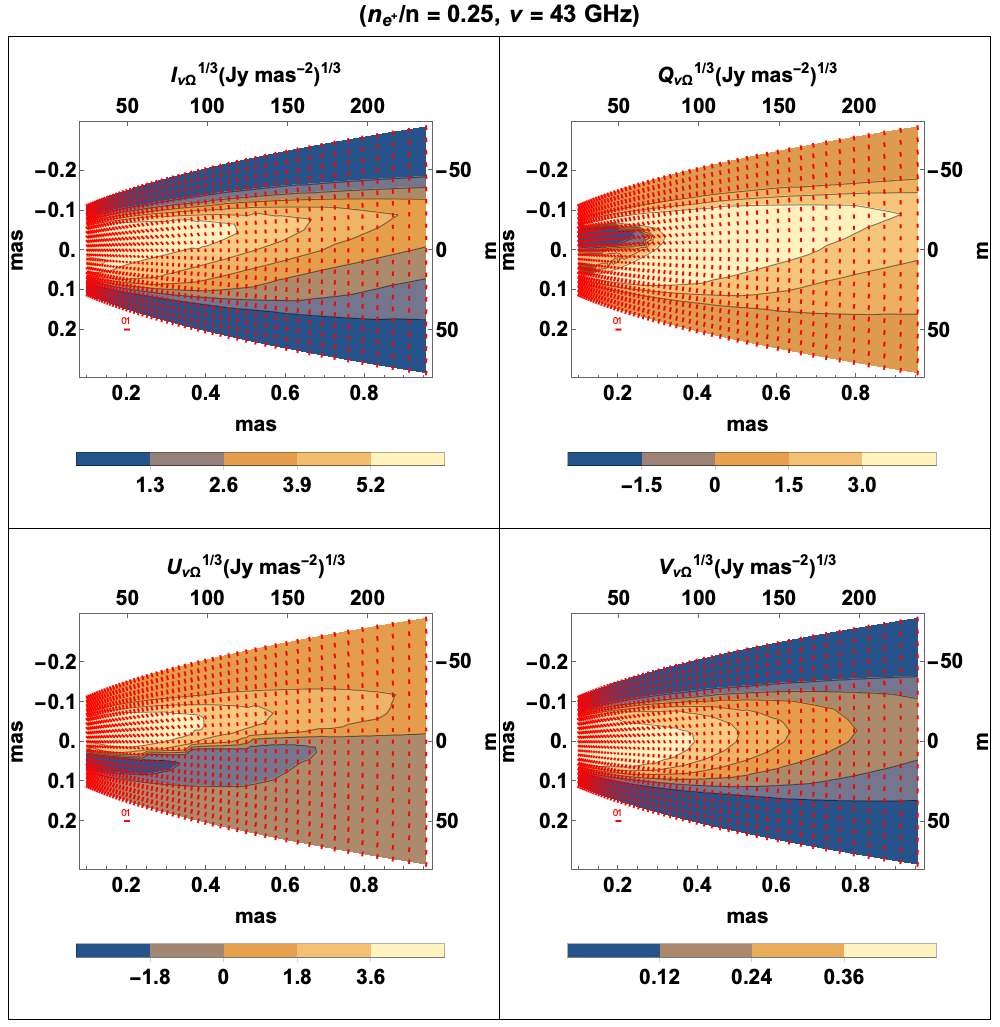} \end{align}  
\begin{align}\nonumber
\includegraphics[height=270pt,width=320pt,trim = 6mm 1mm 0mm 1mm]{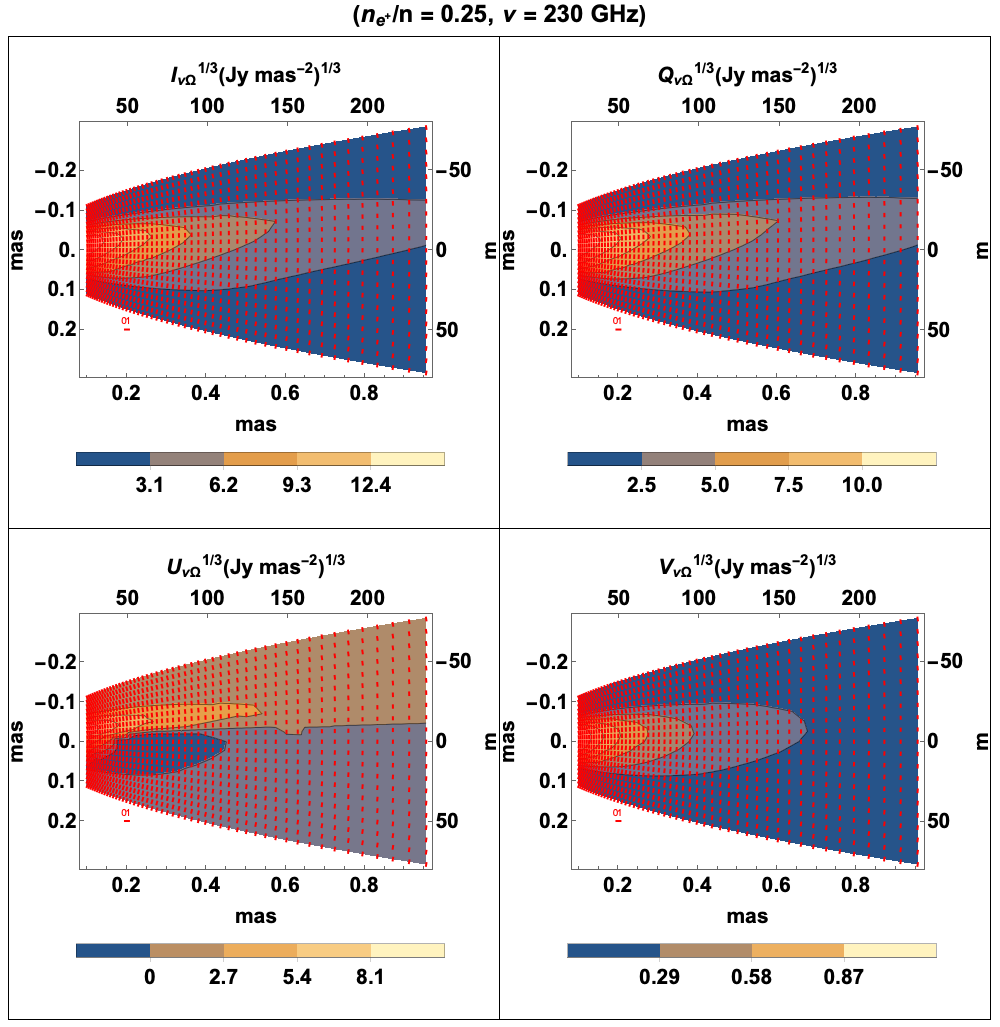}\end{align}
\caption{Stokes map (colors) overlaid with $E$-field vector orientations (red) with $\beta_{e0}=10^{-4}$ at 43 GHz (upper) panel and at 230 GHz (lower) panel. We consider a midway hadronic plasma with $n_{e+}/n = 0.25$ and for  
$\gamma_\mathrm{min}=10$, $\gamma_\mathrm{max}=\infty$.
}\label{StokesMapsGamMin10GamMaxInfNu43GHzBetaE010m4neOnnPt25}
\end{figure}

\subsubsection{
Stokes Map Frequency Comparison for $n_{e^+}/n=0.5$,  $\beta_{e0}=10^{-4}$}
\begin{figure}[H]\nonumber
\begin{align}
\includegraphics[height=270pt,width=320pt,trim = 6mm 1mm 0mm 1mm]{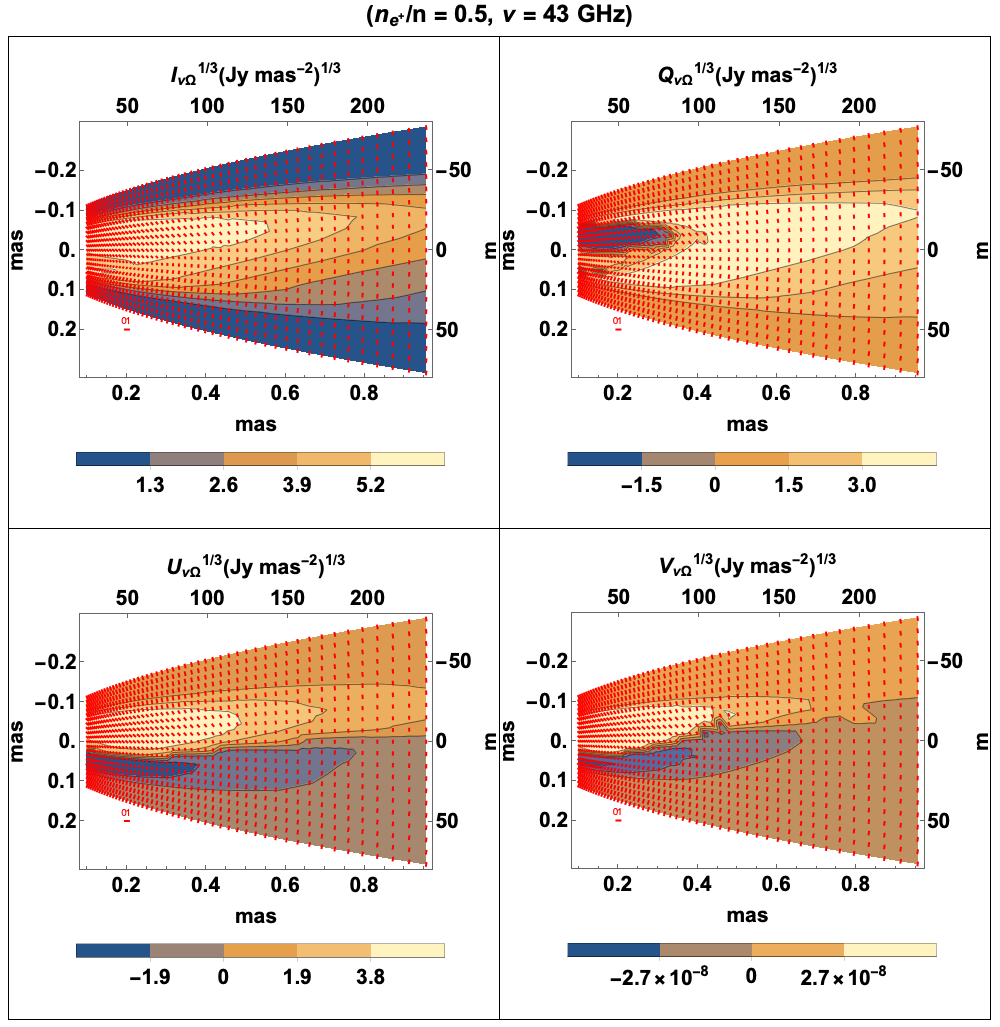} \end{align}  
\begin{align}\nonumber
\includegraphics[height=270pt,width=320pt,trim = 6mm 1mm 0mm 1mm]{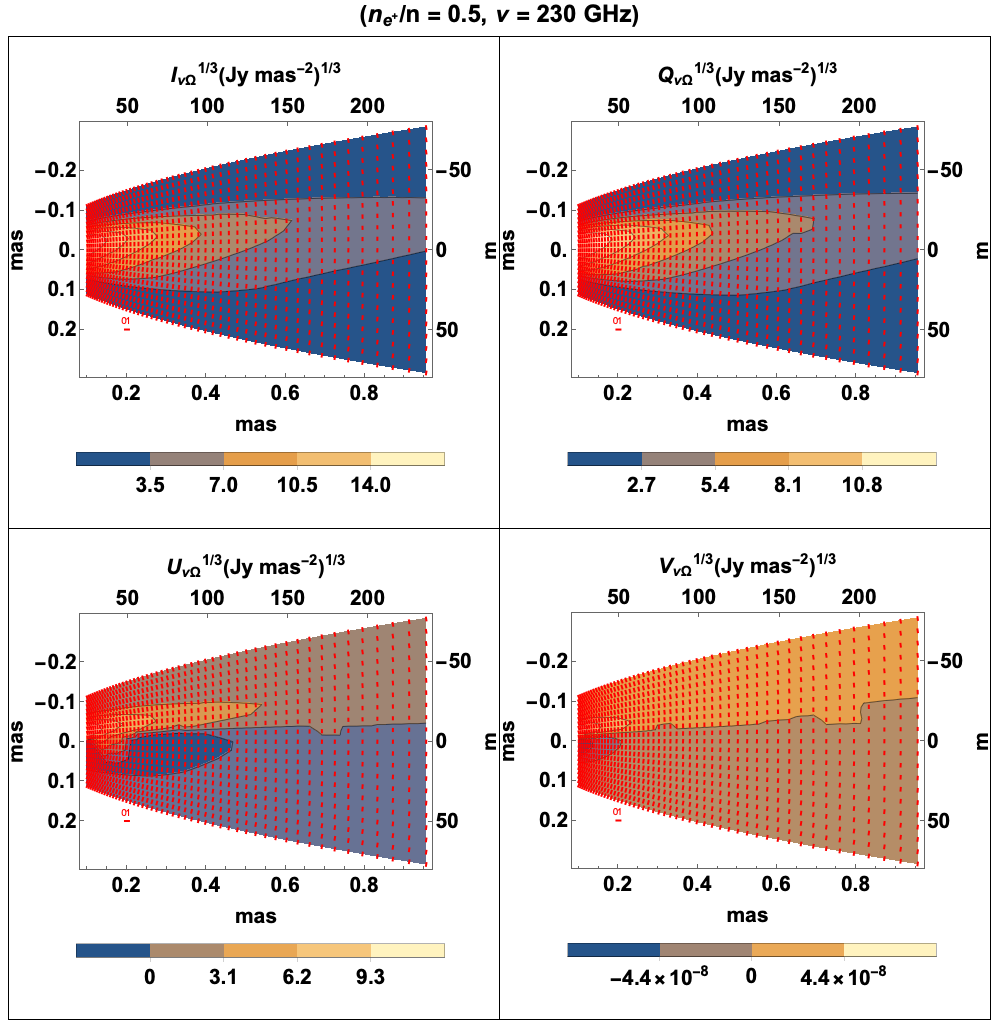}\end{align}
\caption{Stokes map (colors) overlaid with $E$-field vector orientations (red) with $\beta_{e0}=10^{-4}$ at 43 GHz (upper) panel and at 230 GHz (lower) panel. We consider a maximally symmetric $e^{+}e^{-}$ plasma with $n_{e+}/n = 0.5$ and for 
$\gamma_\mathrm{min}=10$, $\gamma_\mathrm{max}=\infty$.
}\label{StokesMapsGamMin10GamMaxInfNu43GHzBetaE010m4neOnnPt5}
\end{figure}

\subsubsection{
Stokes Map Frequency Comparison for $n_{e^+}/n=0.0$,  $\beta_{e0}=10^{-2}$ GHz}
\begin{figure}[H]\nonumber
\begin{align}
\includegraphics[height=270pt,width=320pt,trim = 6mm 1mm 0mm 1mm]{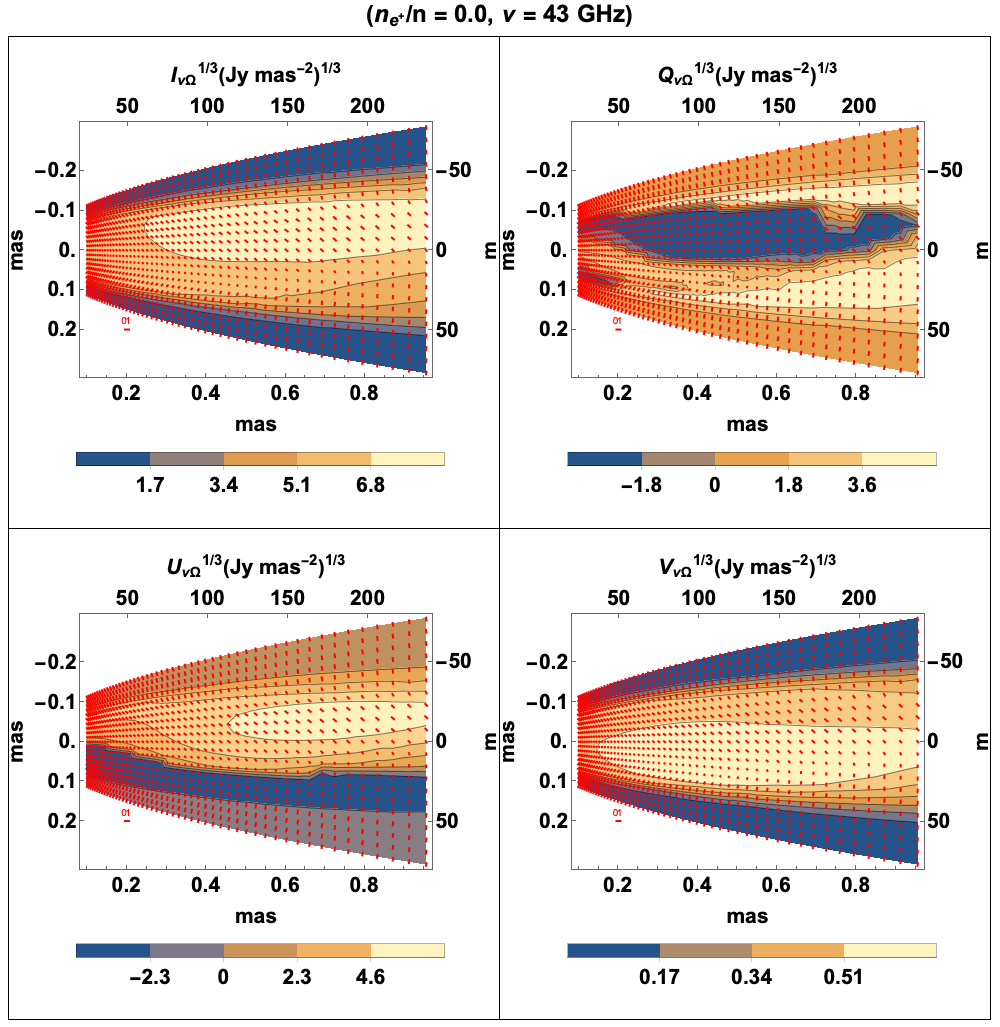} \end{align}  
\begin{align}\nonumber
\includegraphics[height=270pt,width=320pt,trim = 6mm 1mm 0mm 1mm]{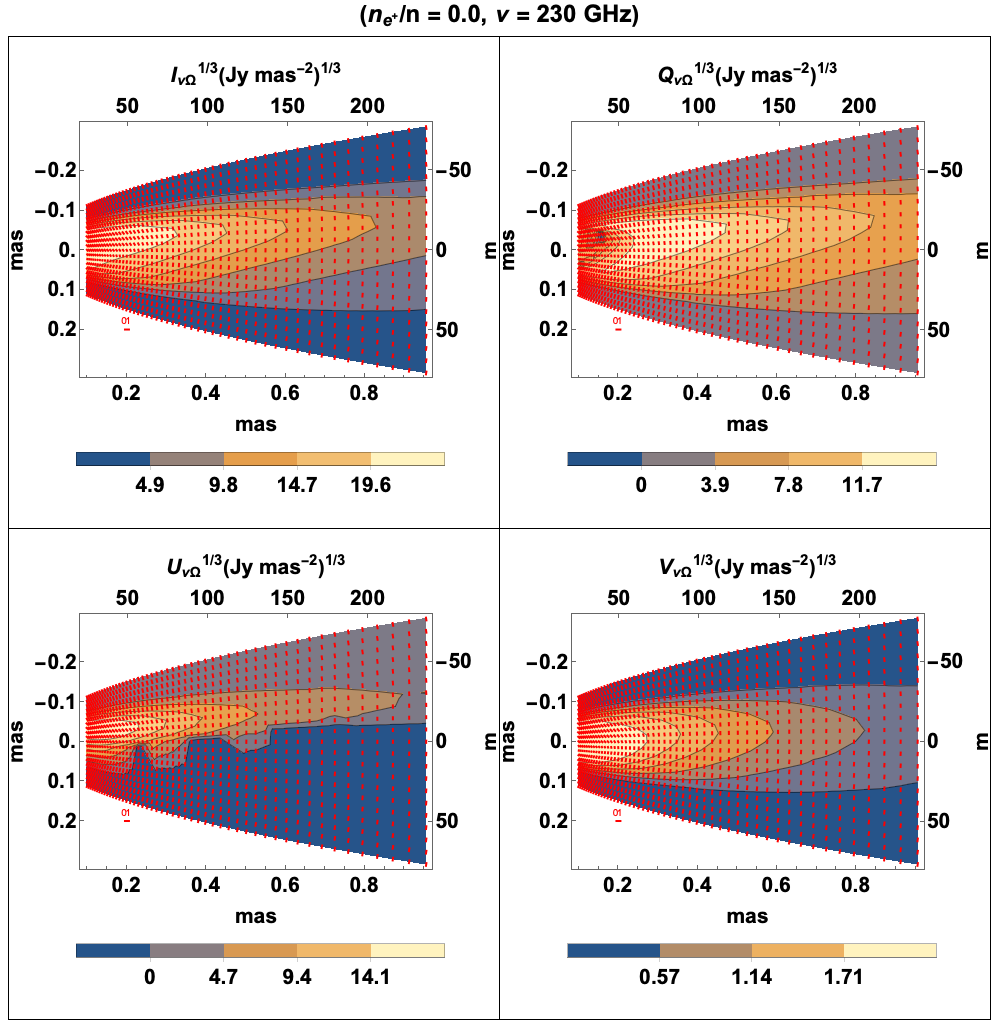}\end{align}
\caption{Stokes map (colors) overlaid with $E$-field vector orientations (red) with $\beta_{e0}=10^{-2}$ at 43 GHz (upper) panel and at 230 GHz (lower) panel. We consider a maximally hadronic plasma with $n_{e+}/n = 0$ and for 
$\gamma_\mathrm{min}=10$, $\gamma_\mathrm{max}=\infty$.
}\label{StokesMapsGamMin10GamMaxInfNu43GHzBetaE010m2neOnnPt0}
\end{figure}

\subsubsection{
Stokes Map Frequency Comparison for $n_{e^+}/n=0.25$, $\beta_{e0}=10^{-2}$}
\begin{figure}[H]\nonumber
\begin{align}
\includegraphics[height=270pt,width=320pt,trim = 6mm 1mm 0mm 1mm]{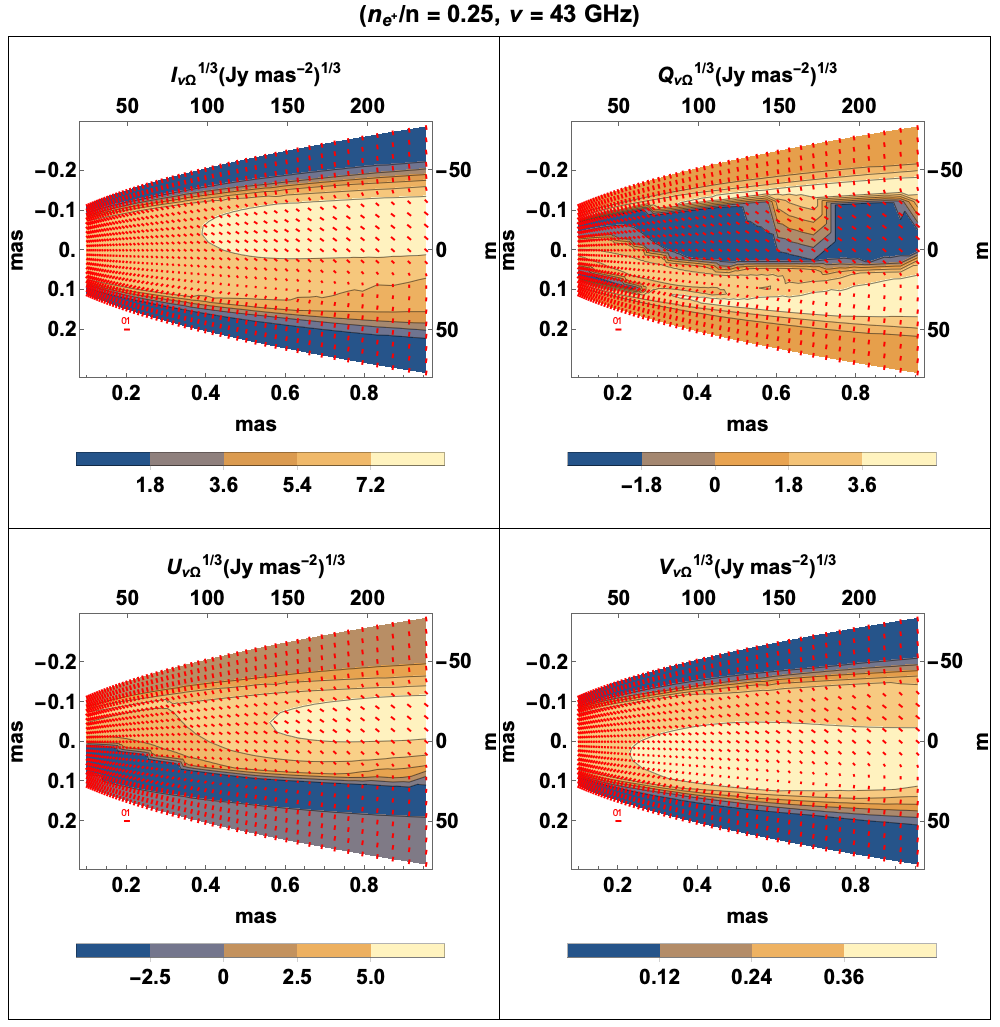} \end{align}  
\begin{align}\nonumber
\includegraphics[height=270pt,width=320pt,trim = 6mm 1mm 0mm 1mm]{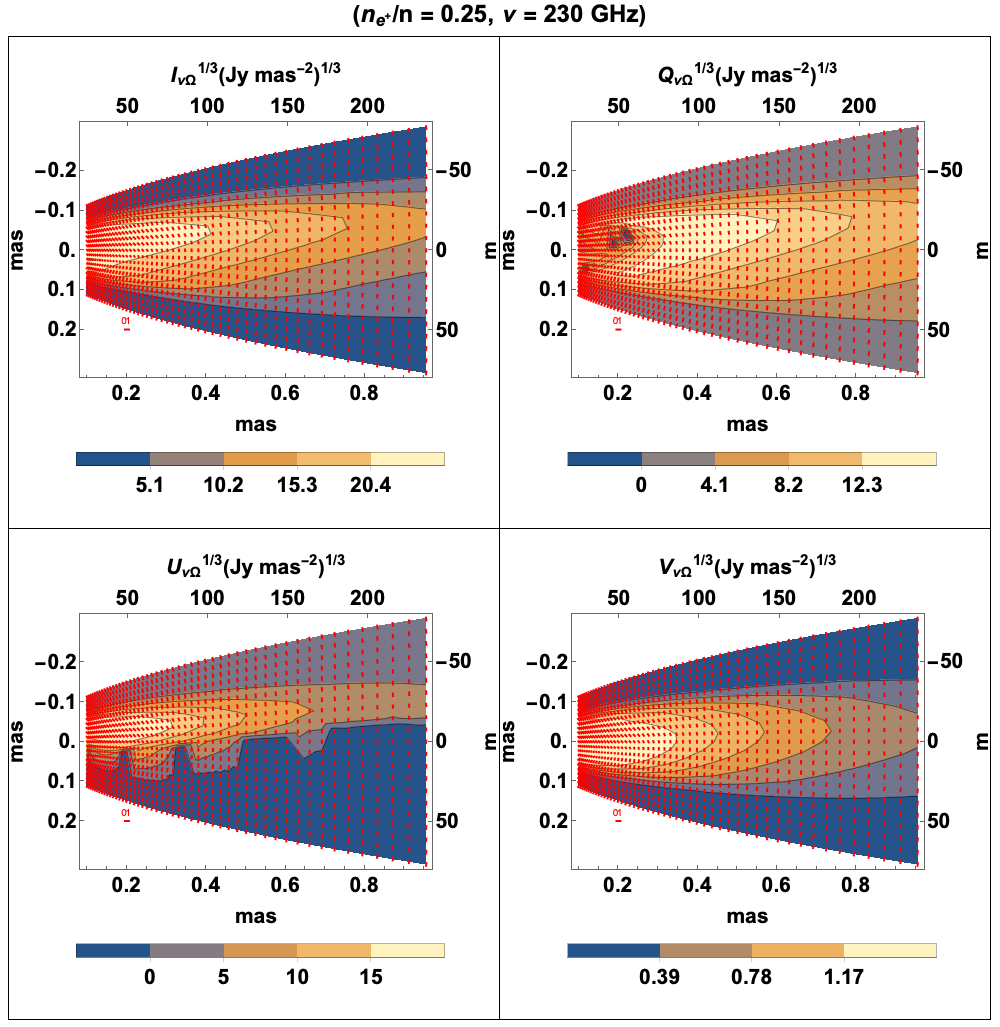}\end{align}
\caption{Stokes map (colors) overlaid with $E$-field vector orientations (red) with $\beta_{e0}=10^{-2}$ at 43 GHz (upper) panel and at 230 GHz (lower) panel. We consider a midway hadronic plasma with $n_{e+}/n = 0.25$ and for 
$\gamma_\mathrm{min}=10$, $\gamma_\mathrm{max}=\infty$.
}\label{StokesMapsGamMin10GamMaxInfNu43GHzBetaE010m2neOnnPt25}
\end{figure}

\subsubsection{
Stokes Map Frequency Comparison for $n_{e^+}/n=0.5$, $\beta_{e0}=10^{-2}$}
\begin{figure}[H]\nonumber
\begin{align}
\includegraphics[height=270pt,width=320pt,trim = 6mm 1mm 0mm 1mm]{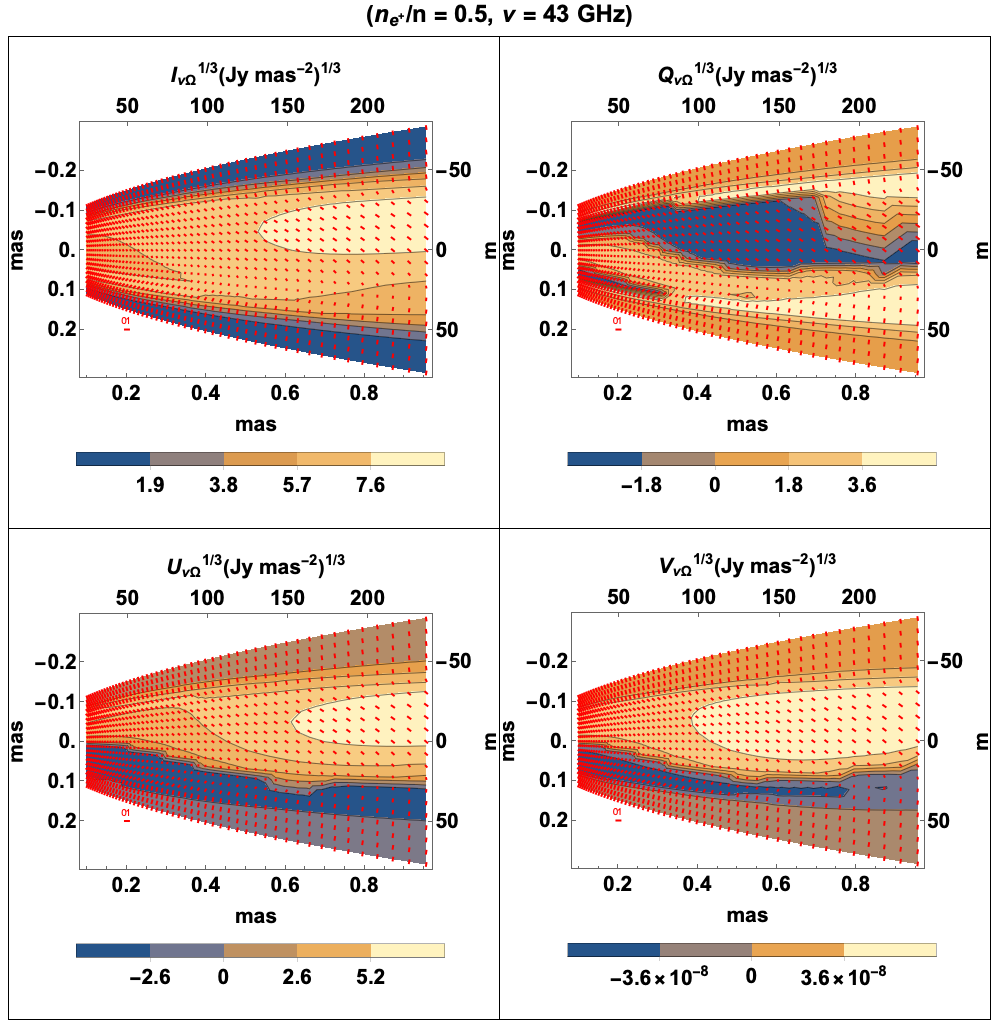} \end{align}  
\begin{align}\nonumber
\includegraphics[height=270pt,width=320pt,trim = 6mm 1mm 0mm 1mm]{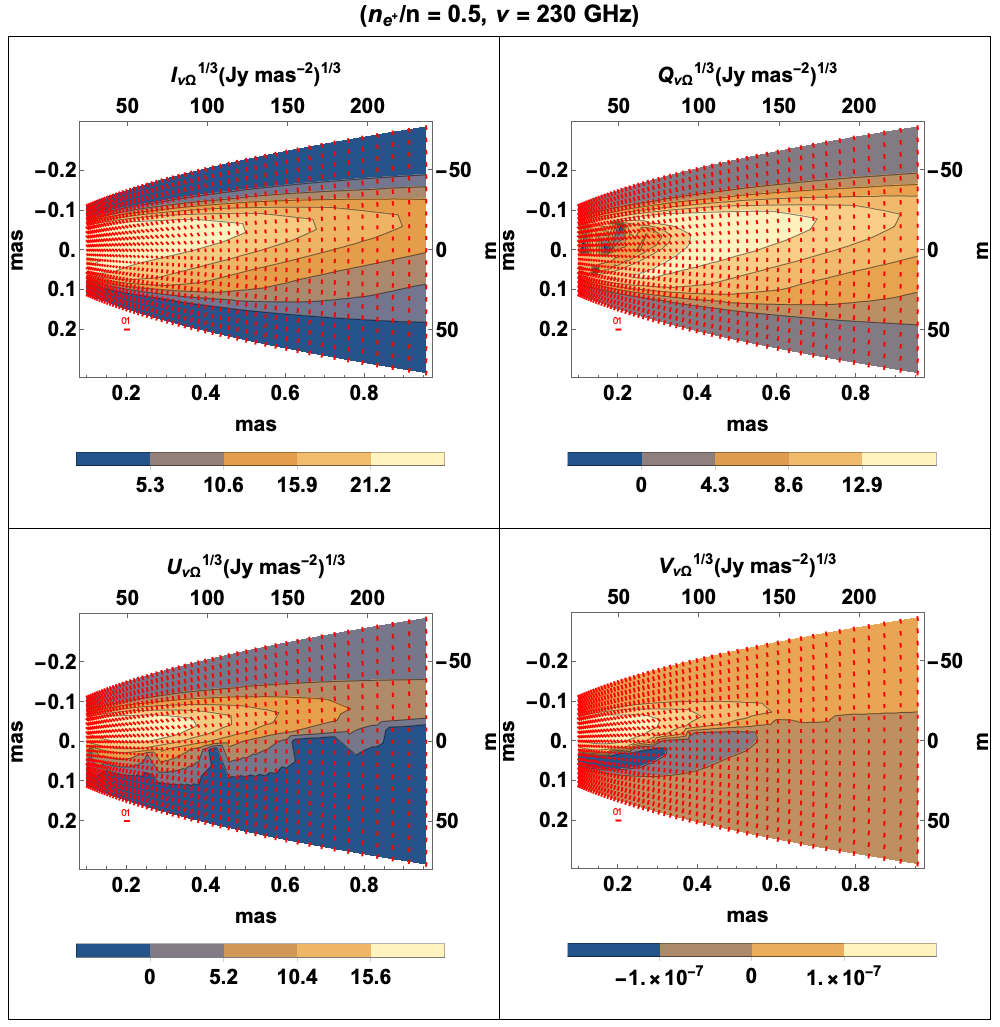}\end{align}
\caption{Stokes map (colors) overlaid with $E$-field vector orientations (red) with $\beta_{e0}=10^{-2}$ at 43 GHz (upper) panel and at 230 GHz (lower) panel. We consider a maximally symmetric $e^+ e^-$ plasma with $n_{e+}/n = 0.5$ and for 
$\gamma_\mathrm{min}=10$, $\gamma_\mathrm{max}=\infty$.
}\label{StokesMapsGamMin10GamMaxInfNu43GHzBetaE010m2neOnnPt5}
\end{figure}
\subsubsection{Autocorrelation Analysis}
\label{AutocorrelationAnalysis}
\begin{figure}[H]\nonumber  
\begin{align}
\includegraphics[height=240pt,width=530pt,trim = 6mm 1mm 0mm 1mm]{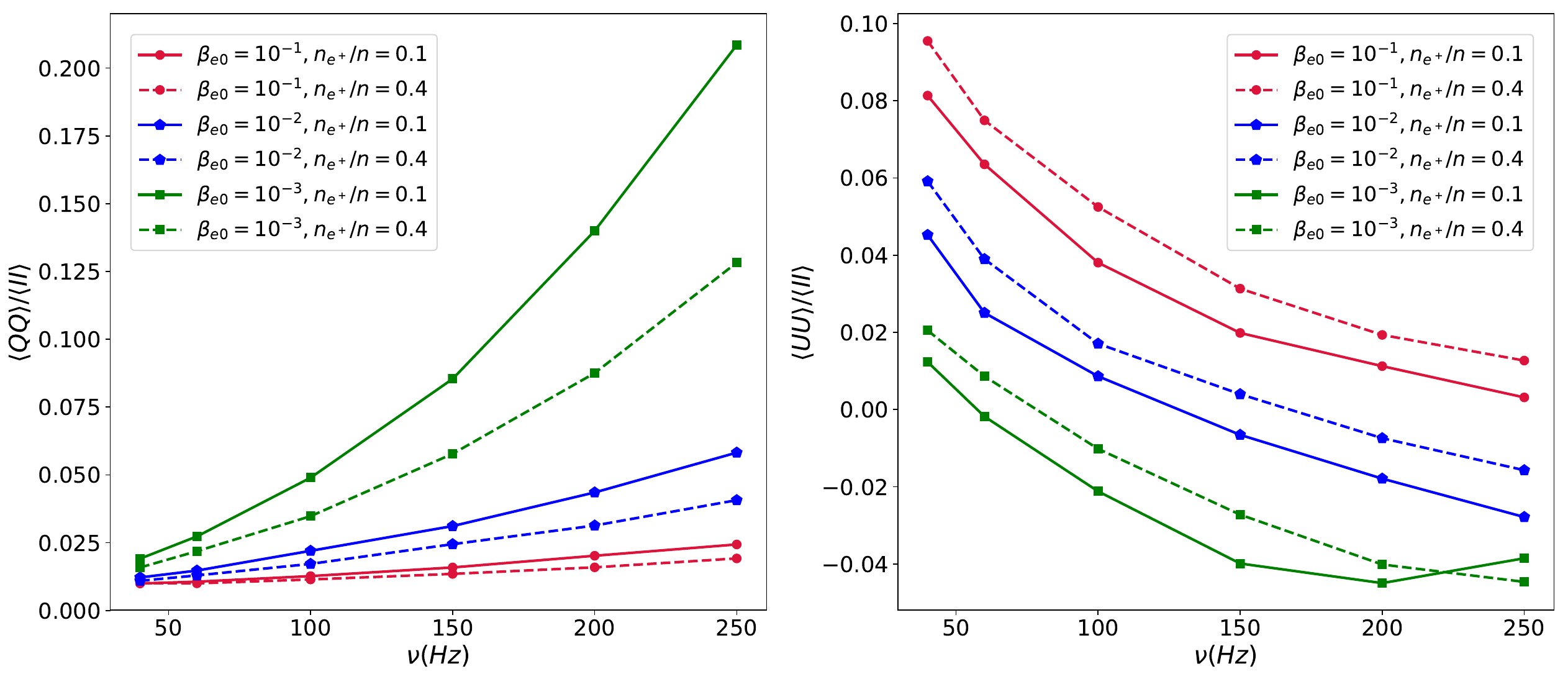
}
\end{align} 
\caption{Autocorrelation ratios $\langle Q Q\rangle/ \langle I I\rangle$  (left) and $\langle U U\rangle/ \langle I I\rangle$  (right) as functions of frequency for different values of $\beta_{e0}= 10^{-3}, 10^{-2}\ \mathrm{and}\ 10^{-1}$ and plasma compositions $n_{e^+}/n = 0.1\ \mathrm{and\ }0.4$. 
}\label{Autocorrelation}
\end{figure}

Next, we calculate the autocorrelation\footnote{Strictly speaking, the autocorrelation of a real valued function $f$, $R_{ff}(\tau)\equiv\int_{-\infty}^\infty f(t)f(t-\tau)dt$, is the integral of the function times the function at a constant lag. Here, we extend the definition to the integral of a function times the function at a multiplicative scaling of the argument. Alternatively, we may use the definition of convolution ($(g*h)(\eta)\equiv\int_{-\infty}^\infty g(\hat{Y})h(\eta-\hat{Y}) d\hat{Y}$) to express $\langle ss \rangle = \int_{-\infty}^\infty (s*s)(0) d\hat{X}$, where $\hat{X}=X$ if $|X|<X_\mathrm{max}$, 0 otherwise, and analogously for $Y$,
as the integral of the convolution of $s$ with itself centered at $Y=0$.
} functions of Stokes parameters, where for a parabolic jet propagating in the $Y$ direction, this is defined as: 

\begin{equation}
\label{auto-Correlation-X}
\langle ss
\rangle = \int_{X_\mathrm{min}}^{X_\mathrm{max}} dX \int_0^{Y_\mathrm{max}} dY ~ 
s(X,Y) 
s(X, -Y) ~~~,~~~  s \in \{I, Q, U, V\}.
\end{equation}
where the integrals are taken over the entire map for each Stokes parameter, as presented in Figures (\ref{StokesMapsGamMin10GamMaxInfNu43GHzBetaE010m4neOnnPt0}-\ref{StokesMapsGamMin10GamMaxInfNu43GHzBetaE010m2neOnnPt5}). We start with computing the autocorrelations 
of $Q$ and $U$ as functions of frequency. 
The autocorrelation functions are very useful parameters to exhibit the symmetry/antisymmetry patterns of the Stokes $Q$ and $U$ parameters. In the following, we normalize these autocorrelation functions to that of the Stokes parameter $I$.
%
%
%
Figure \ref{Autocorrelation} presents  $\langle Q Q\rangle/\langle I I\rangle $ and $\langle U U \rangle/ \langle I I \rangle$ for different values of $\beta_{e0}$ as well as different plasma compositions, where again we adopt $\gamma_\mathrm{min}$ = $10$ and $\gamma_\mathrm{max}$ = $\infty$. 

The symmetric/antisymmetric patterns of the Stokes maps presented in Section \ref{SymProperties} are replicated in the autocorrelation functions through a sign difference between the $Q$ and $U$ autocorrelations in Figure \ref{Autocorrelation}. This more readily achieved for lower values of $\beta_{e0}$, smaller values of $n_{e^+}/n$ and for higher frequencies. The positive autocorrelation for $Q$ in Fig. \ref{Autocorrelation} intuitively reflects the symmetry in its maps about $Y=0$. At higher frequencies and for smaller values of $\beta_{e0}$, $U$ displays its anticipated anti-symmetry (note this is the orthogonal polarization to $Q$). However, at lower frequencies, the $U$ map contours bow downward in a manner providing large positive contributions to the autocorrelation for low $|Y|$. This leads to a net positive autocorrelation for $U$ in these cases. 



Figure \ref{Autocorrelation} gives us the following predictions:

$\bullet$ Increasing the frequency increases (decreases) the value of $\langle Q Q\rangle/ \langle I I\rangle$ ($\langle U U \rangle/ \langle I I\rangle$), respectively. 

$\bullet$ Larger values of $\beta_{e0}$ enhance the auto-correlation of Stokes $I$ faster than that of $Q$. Therefore the 
quantity $\langle Q Q\rangle/ \langle I I\rangle$ decreases with increasing $\beta_{e0}$.

$\bullet$ On the other hand, increasing $\beta_{e0}$ washes out the antisymmetric patterns in the normalized $U$ maps, and does so more rapidly at lower frequencies. 
There is a characteristic frequency, $\nu^*$,  at which the autocorrelation of $U$ is positive for $\nu < \nu^*$ and is negative for $\nu > \nu^*$. Increasing the value of $\beta_{e0}$ as well as $n_{e^+}/n$ increases this frequency.

Owing to these important properties, we claim that multi-wavelength observations of the Stokes parameters can be utilized to break the degeneracy between parameters $\beta_{e0}$ and $n_{e^+}/n$ and shed more light on the composition of the plasma. We note that for a given AGN jet, these findings are sensitive to whether $Q$ and $U$ are defined relative to the observer plane axes in Fig. \ref{PolarizationAxes}, where the forward jet projection expands along the positive $Y$ axis. Across an ensemble of jets randomly oriented on the observer plane, we expect $Q$ and $U$ to each take on bilaterally symmetric and antisymmetric properties, but for each jet, the $Q$-$U$ phase difference ensures that the symmetry properties of $Q$ are complementary to those of $U$ at any observer time. 

\begin{figure}[H]\nonumber  
\begin{align}
\includegraphics[height=300pt,width=340pt,trim = 6mm 1mm 0mm 1mm]{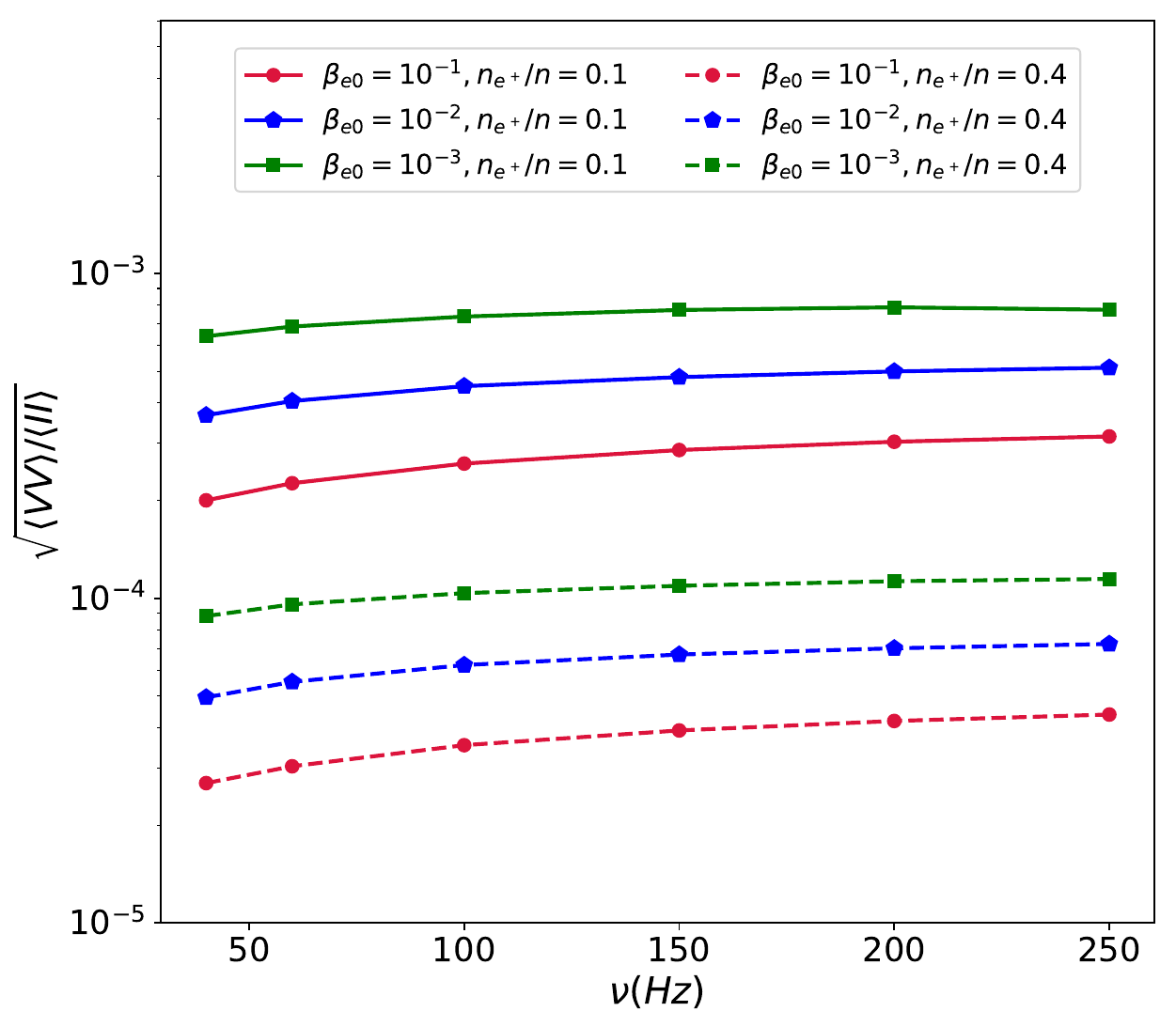
}
\end{align} 
\caption{Autocorrelation ratio $\sqrt{\langle VV \rangle /\langle II  \rangle}$ as a function of frequency for different values of $\beta_{e0}= 10^{-3}, 10^{-2}\ \mathrm{and}\ 10^{-1}$ and plasma compositions $n_{e^+}/n = 0.1\ \mathrm{and\ }0.4$. 
}\label{Autocorrelation-VV}
\end{figure}

Finally, we calculate the autocorrelation of $V$, normalize it to $\langle I I \rangle$ and take the square root.\footnote{All values of $\langle VV \rangle$ computed in this analysis are positive.} Figure \ref{Autocorrelation-VV} presents  $\sqrt{\langle V V \rangle /\langle II  \rangle}$ as a function of the frequency  for different values of $\beta_{e0}$ as well as different plasma compositions.
\noindent
Unlike the normalized autocorrelation functions of $Q$ and $U$, $\sqrt{\langle V V \rangle /\langle II  \rangle}$ is more sensitive to the plasma composition than to the value of the $\beta_{e0}$. This means we can start to break the aforementioned degeneracy between physical parameters if we combine the analyses of linear and circular polarizations. The observed frequency dependence in $\sqrt{\langle V V \rangle /\langle II  \rangle}$ can be used to predict plasma composition while  
those in $\langle QQ\rangle/\langle II  \rangle$ and $\langle UU\rangle/\langle II  \rangle$ 
are used to determine $\beta_{e0}$.
Taking Figs. \ref{Autocorrelation} and  \ref{Autocorrelation-VV} together, it is clear that the spectrum of $\sqrt{\langle V V \rangle /\langle I I \rangle}$ is flatter than that of the more anti-symmetric linear polarization ($U$) across parameter space and flatter than that of the more symmetric linear polarization ($Q$), particularly for lower $\beta_{e0}$, 
making the circular polarization autocorrelation a more universal quantity.

\subsection{Degree of Polarization}
 
Polarization has been robustly detected in many AGN. M87 has been observed at 86 GHz by the VLBA and Green Bank Telescope to be up to $20\%$ polarized near the core \citep{Hada2016}. The preponderance of this is assumed to be linear polarization from jet regions with highly ordered magnetic fields.  Percent-level degrees of circular polarization have been detected in Sgr A* ($1.2\%\pm0.3\%$ at 230 GHz and $1.6\%\pm0.3\%$ at 345 GHz \citep{Munoz2012}) by the Submillimeter Array (SMA) to tenths-of-a-percent precision; however, the determination of the source's intrinsic circular polarization is more challenging due to the low magnitude of circular polarization and is further confounded by Faraday conversion. The SPRITE proposal at OVRO plans to discriminate among observations of the degree of circular polarization at the $10^{-3}$ level \citep{Blandford2019}; however, a modeling complication is that the emissivity is sensitive to the pitch angle distribution of the emitting particles (though it is worth calculating assuming an isotropic distribution). We now present degree of circular polarization maps for our semi-analytic model, with signatures that promise to distinguish observed plasmas with different compositions.

 \subsubsection{
Degree of Circular Polarization
}

\begin{figure}
\nonumber
\begin{align}
\includegraphics[height=170pt,width=240pt,trim = 6mm 1mm 0mm 1mm]{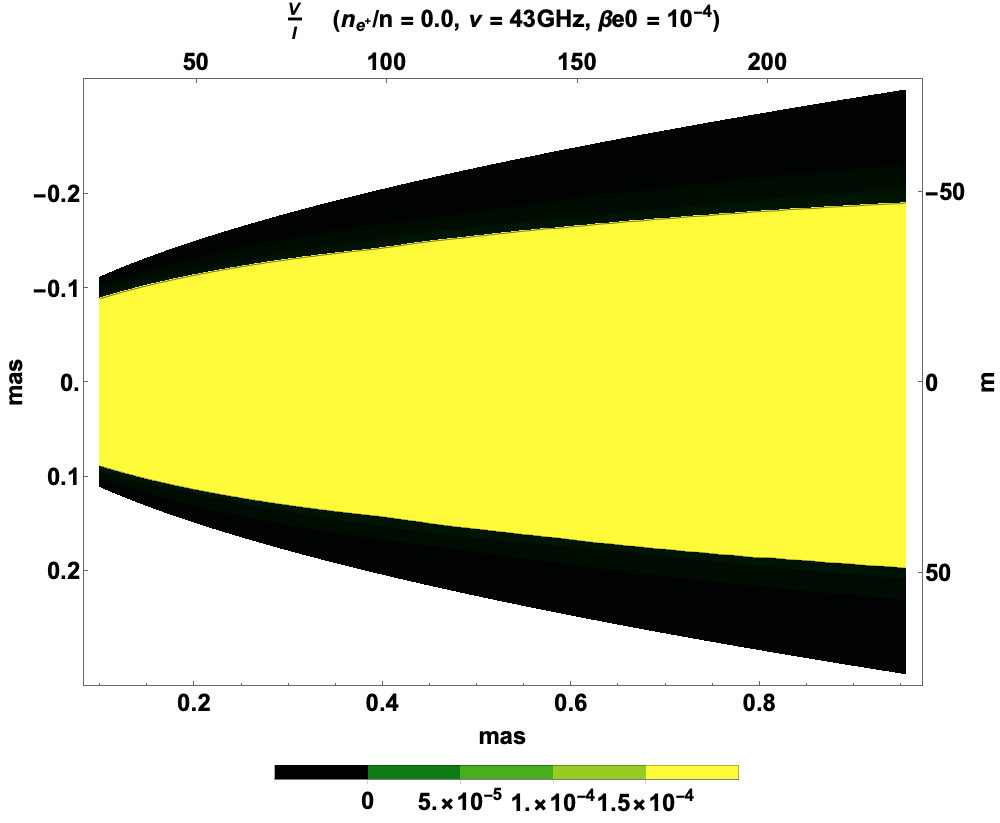}\ \ \   &\hspace{1cm}\includegraphics[height=170pt,width=240pt,trim = 6mm 1mm 0mm 1mm]{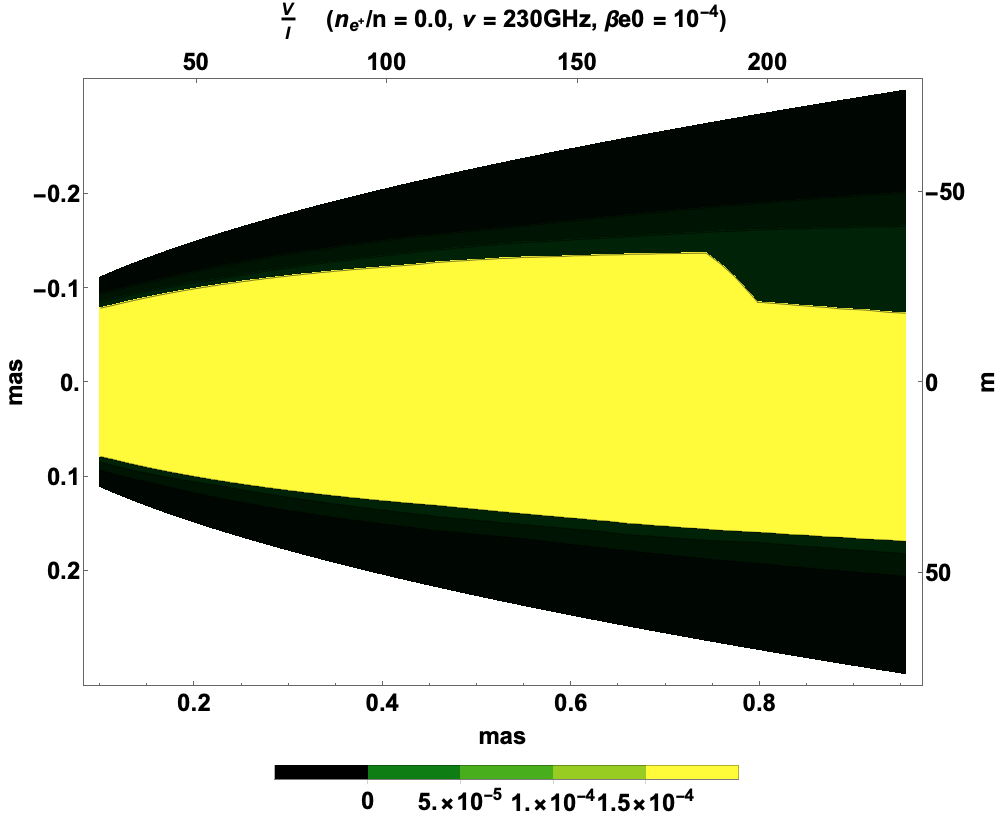}
\end{align}   
\begin{align}
\includegraphics[height=170pt,width=240pt,trim = 6mm 1mm 0mm 1mm]{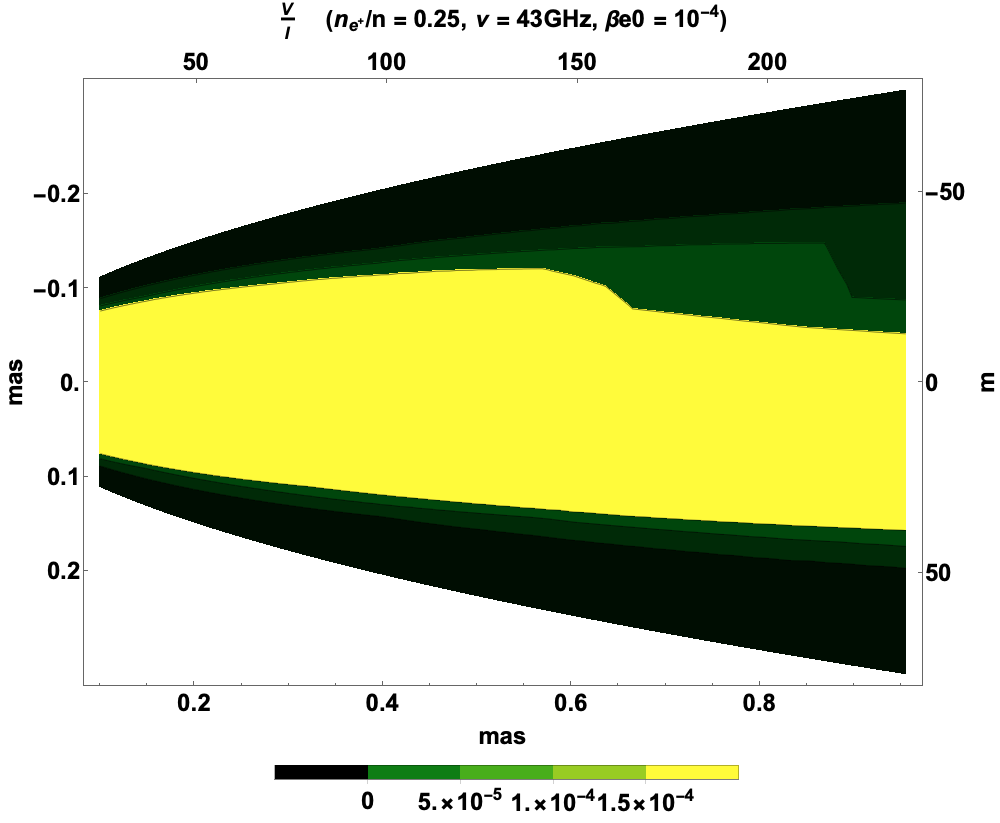}\ \ \  &\hspace{1cm}\includegraphics[height=170pt,width=240pt,trim = 6mm 1mm 0mm 1mm]{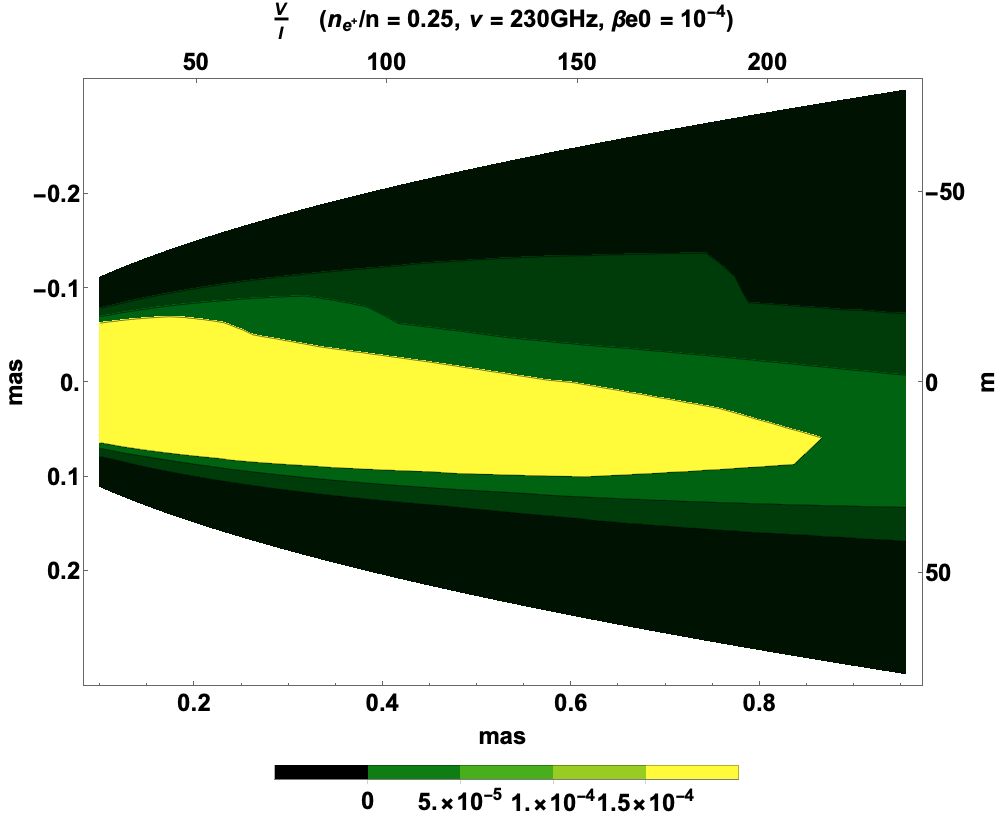}
\end{align} 
 \begin{align}\nonumber
 \includegraphics[height=170pt,width=240pt,trim = 6mm 1mm 0mm 1mm]{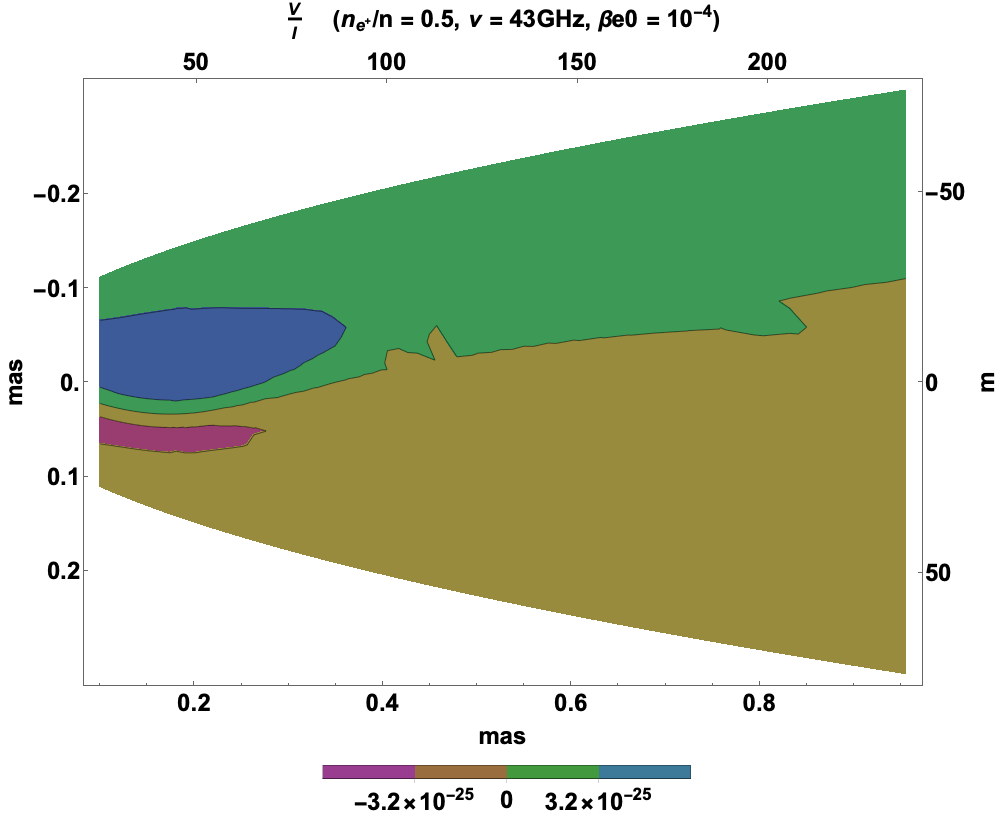}\ \ \  &\hspace{1cm}\includegraphics[height=170pt,width=240pt,trim = 6mm 1mm 0mm 1mm]{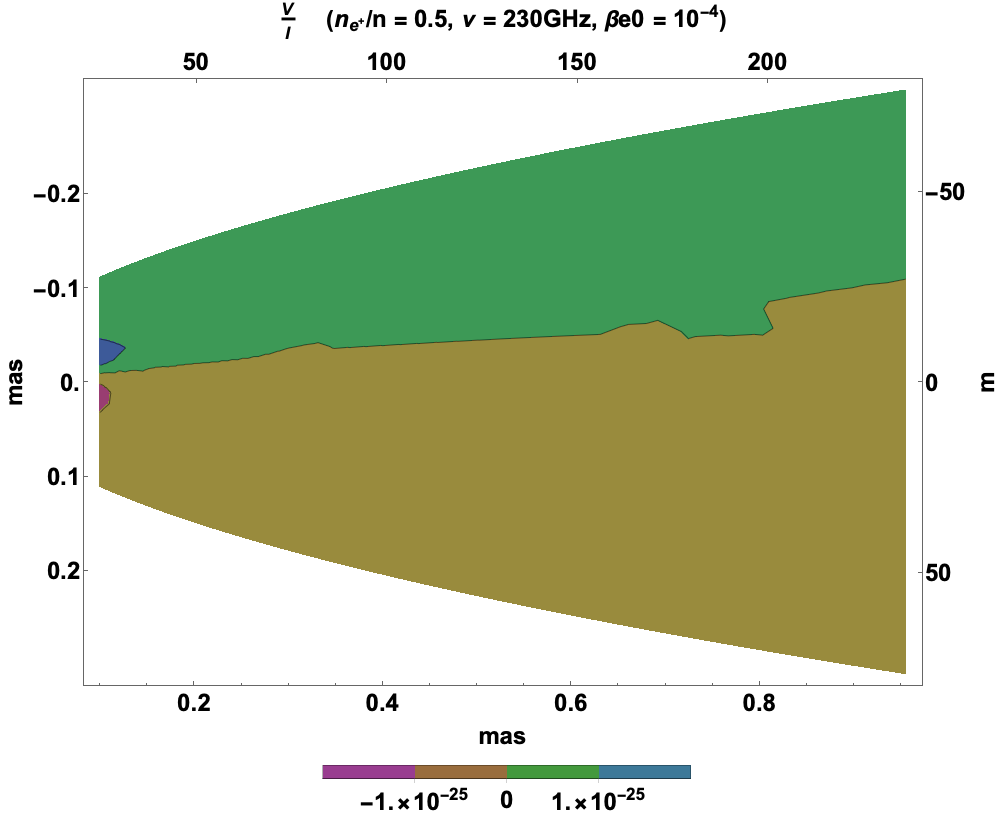}
 \end{align}
\caption{Degree of circular polarization $V/I$ at 43 GHz (left) and 230 GHz (right) for maximally ionic plasma $n_{e^+}=0,\ n_i=0.5$ (top), fiducial model plasma $n_{e+}=n/4=n_i$  and a maximally leptonic plasma $n_{e^+}=n_{e^-}=n/2,\ n_i=0$ (bottom) for $\beta_{e0}=10^{-4}$, $\gamma_\mathrm{min}=10$ and $\gamma_\mathrm{max}=\infty$. 
}\label{DegreeOfVPol43And230GHz} 
\end{figure}

In our models, the degree of circular polarization vanishes for pure $e^-e^+$ plasma (aside from minor Faraday conversion effects) due to nearly canceling contributions from electrons and positrons. This is borne out in Fig. \ref{DegreeOfVPol43And230GHz}, where different legends are used for the lower plots in order to accommodate the tens of orders of magnitude suppression for the $n_{e^+}/n=0.5=n_{e^-}/n$ case. By contrast, plasmas with at least half of positive charge carriers ionic for $\beta_{e0}=10^{-4}$ (cf. Table \ref{CircPolDeg}) have  $V/I$ at least a few $10^{-5}$ over most of the observer plane, with maximum/mean values  
%
$8.0\times 10^{-4}/1.4\times 10^{-4}$ for $(n_{e^+}/n=0.25,\nu=43$ GHz), $5.7\times 10^{-4}/7.0\times 10^{-5}$ for $(n_{e^+}/n=0.25,\nu=230$ GHz), $2.7\times 10^{-3}/4.6\times 10^{-4}$ for $(n_{e^+}/n=0.0,\nu=43$ GHz) and $1.6\times 10^{-3}/2.1\times 10^{-4}$ for $(n_{e^+}/n=0.0,\nu=230$ GHz) (cf. Table 1). For maximally hadronic plasma, the tenths of a percent circular 
levels predicted by our models are currently measurable, e.g., by SMA and EHT, or easily within the realm of possibility for near future instruments.

\begin{table}[H]
	\centering
	\caption{Mean and maximum degree of circular polarization for $\beta_{e0}=10^{-4}$ models.
    }
	\label{CircPolDeg}
	\begin{tabular}{lccccccr} 
		\hline
		Model ($n_{e^+}/n,\nu$) & 
        && Mean $V/I$ && 
        & 
        Max $V/I$ &  \\
		\hline
        (0.0, 43 GHz)& 
        &&   
        $4.6 \times 10^{-4}$
        && 
        & $2.7 \times 10^{-3}$
        &\\
       (0.0, 230 GHz)&  
       && $2.1 \times 10^{-4}$ 
       && 
       &  $1.6 \times 10^{-3}$ 
       &\\
       (0.25, 43 GHz)&  
       & 
       &   $1.4 \times 10^{-4}$ 
       &
       && $8 \times 10^{-4}$ 
       & \\
       (0.25, 230 GHz) & 
       &&  $5.7 \times 10^{-4}$ 
       &  
       & 
       &  $7.0 \times 10^{-5}$ 
       &\\
       (0.5, 43 GHz)&  
       && $3 \times 10^{-26}$
       && 
       & $6 \times 10^{-25}$
       &\\
       (0.5, 230 GHz)&  
       && $2\times 10^{-25}$
       && 
       & $-2 \times 10^{-28}$
       &\\ 
        \hline
	\end{tabular}
\end{table}

\section{Conclusions}
\label{conc}

In this work, we address the long standing mystery of the plasma composition of relativistic jets not only from the standpoint of intensity, but rather through the lens of all of four Stokes parameters. To this end, we implement a simple equipartition-based emission model in which the particle pressure is taken to be fixed fractions of the magnetic pressure in a self-similar, stationary, axisymmetric M87-based jet model with electrons and varying positive ion versus positron content. Visualizing polarization as Stokes maps as opposed to $E$-field directions overlaying intensity maps confers the advantage of enabling symmetries to manifest themselves. Among the observational signatures predicted are slight bilateral asymmetry in $I$ and one independent linear polarization, and bilateral anti-symmetry in the other. The clearest observable is an enhancement of circular polarization with increasing parameter $\beta_{e0}$ and ion content $n_i/n$. Thus, we have the immediate mm to sub-mm observational prospect of inferring $e^-e^+$ plasmas near equipartition ($\beta_{e0}\lesssim 0$) if the circular polarization degree is vanishing, or that a magnetically dominated plasma ($\beta_{e0}<< 0$) has mostly $p$ positive charge carriers if the degree of polarization is measurable ($\gtrsim 0.1\%$). 

We also considered the autocorrelation functions of the Stokes parameters and normalized them to that of the intensity. The normalized autocorrelations of Stokes $Q$ and $U$ increase and decrease with the frequency, respectively. Our computations show that both of these parameters exhibit more sensitivity to the value of $\beta_{e0}$ than the plasma composition. The symmetric/antisymmetric pattern of Stokes $Q$ and $U$ correspondingly depend on the value of $\beta_{e0}$. Increasing $\beta_{e0}$ washes out the antisymmetric patterns in the $U$ maps up to a characteristic frequency, $\nu^*$, at which the autocorrelation of $U$ changes sign. The antisymmetric patterns then show up for higher frequencies. Increasing $n_{e^+}/n$ increases this characteristic frequency. Larger values of $\beta_{e0}$ enhance the autocorrelation of Stokes $I$ more quickly than that of Stokes $Q$, thus decreasing the magnitude of $\langle Q Q\rangle/ \langle I I\rangle$ with increasing $\beta_{e0}$; higher $n_{e^+}/n$ tends to flatten $\langle Q Q\rangle/ \langle I I\rangle(\omega)$. We may use these as probes indicating that rapid increase in the autocorrelation of a normalized linear polarization 
may favor lower $\beta_{e0}$ or higher ionic content. 
On the other hand, we find that 
$\sqrt{\langle V V \rangle /\langle II  \rangle}$ is universally flat with respect to the frequency and shows more sensitivity to the plasma composition than the value of $\beta_{e0}$. 
Thus, the simultaneous consideration of $Q, U$ and $V$ can break the degeneracy between the physical parameters $n_{e^+}/n$ and $\beta_{e0}$ and give us novel information about the underlying physics. 

We only used a simple equipartition-inspired emission prescription 
in our semi-analytic model in this work. In the future, we plan to consider other 
emission models based on velocity shear and jet current density \citep{Anantua2018}. Some models may affect intensity and polarization differently due to different dependencies of the magnetic field strength $b$ on the strengths of magnetic field components $B_{||}$ and $B_\perp$ with respect to the observer direction. Our model was tuned to the viewing angle of the M87 jet, though sources with smaller viewing angles are subject to potentially substantial additional Doppler boosting effects $\propto \mathcal{D}^4$. We could also extend the model using self-similarity to compare a large scale synthetic jet to observations of jets as cosmic ray accelerators \citep{Fowler2019}; or change our computational apparatus to a polarized Kerr radiative transfer routine such as GRTRANS \citep{2016MNRAS.462..115D} in order to ray trace our model near the black hole.
\section*{ACKNOWLEDGMENTS}

R. A. and R.E. acknowledge support by the Institute for Theory and Computation at the Center for Astrophysics $|$ Harvard and Smithsonian. This work was supported in part by the Black Hole Initiative at Harvard University, which is funded by grants from the John Templeton Foundation and the Gordon and Betty Moore Foundation. 

\end{document}